\documentclass[twocolumn,tighten]{aastex631}
\usepackage{amssymb, amsmath,framed}
\usepackage{hyperref}
\usepackage[utf8]{inputenc}
\DeclareUnicodeCharacter{200B}{}  
\DeclareUnicodeCharacter{00A0}{~} 

\usepackage{bm}
\expandafter\ifx\csname package@font\endcsname\relax\else
 \expandafter\expandafter
 \expandafter\usepackage
 \expandafter\expandafter
 \expandafter{\csname package@font\endcsname}
\fi
\def\bq{\begin{equation}}
\def\eq{\end{equation}}
\def\bqy{\begin{eqnarray}}
\def\eqy{\end{eqnarray}}

\def\p{\partial}

\def\p{\partial}

\def\R{\mathbb{R}}

 \def\q#1#2{q^{\hspace{1pt}#1}_{\,,\hspace{1pt}#2}}
 \def\a#1#2{a^{\hspace{1pt}#1}_{\,,\hspace{1pt}#2}}
 \def\d#1#2{\delta^{\hspace{1pt}#1}_{\,,\hspace{1pt}#2}}

\begin{document}
\title{\large{Neural Network Construction of the Equation of State from Relativistic \textit{ab initio} Calculations}}


\author{Kangmin Chen}
\affiliation{School of Physics and Mechatronic Engineering, Guizhou Minzu University, Guiyang 550025, China}

\author{Xiaoying Qu}
\affiliation{School of Physics and Mechatronic Engineering, Guizhou Minzu University, Guiyang 550025, China, xiaoyingqu@126.com}

\author{Hui Tong}
\affiliation{Helmholtz-Institut f\"{u}r Strahlen- und Kernphysik and Bethe Center for Theoretical Physics, Universit\"{a}t Bonn, Bonn D-53115, Germany, tong@hiskp.uni-bonn.de}

\author{Sibo Wang}
\affiliation{Department of Physics and Chongqing Key Laboratory for Strongly Coupled Physics, Chongqing University, Chongqing 401331, China, sbwang@cqu.edu.cn}
\affiliation{Department of Physics, Graduate School of Science, The University of Tokyo, Tokyo 113-0033, Japan}

\author{Yangyang Yu}
\affiliation{School of Physics and Mechatronic Engineering, Guizhou Minzu University, Guiyang 550025, China}

\begin{abstract}
Constraining the nuclear matter equation of state (EOS) beyond saturation density is a central goal of nuclear physics and astrophysics.
While the relativistic Brueckner-Hartree-Fock (RBHF) theory, an \textit{ab initio,} non-perturbative nuclear many-body theory starting from realistic interactions, accurately describes nuclear matter properties near the saturation density $\rho_0 \approx 0.16$~fm$^{-3}$, its applicability is currently limited to densities up to $3\rho_0$, necessitating a reliable extrapolation to higher densities.
In this work, we employ supervised machine learning to train thousands of fully connected neural networks on low-density RBHF data.
By enforcing thermodynamic consistency and smoothness, we finally select a subset of 264 optimal models. These models employ the Swish activation function, which we identify as the most reliable choice for stable extrapolation after extensive testing and comparison. Using these models to extend the EOS over the full density range, we obtain the nuclear matter symmetry energy and then compute the neutron star mass-radius relation and tidal deformability, which are in a great harmony with current astronomical observations. 
The corresponding extrapolation uncertainty originates from the combined contributions of both the 264 optimal models and the linear regression on  nuclear matter EOS, yielding a symmetry energy of $E\mathrm{_{sym}(5\rho_0)=136.0 \pm 52.8 ~MeV}$, a pressure of $P(5\rho_0) = 346.3 \pm 97.4 ~\mathrm{MeV/fm^{3}}$, a maximum neutron star mass of $M\mathrm{_{max}=2.18 \pm 0.18}~M_{\odot}$, and a tidal deformability of $\Lambda_{1.4M_\odot} = 532 \pm 34$.
This work establishes a general and data-driven framework to explore dense matter EOS by integrating \textit{ab initio} calculations with modern machine learning techniques.
\end{abstract}

\section{Introduction}\label{SecIntro}

Understanding the equation of state (EOS) of the dense, neutron-rich matter from the fundamental interactions among its constituents remains one of the major challenges in nuclear physics and astrophysics  
~\citep{Pawel_2002_Science,F.Weber_2007_PPNP,Baldo_2016_PPNP,Lattimer_2016_PR,Li_2018_PPNP,Oertel_2017_RMP,Tim.Dietrich_2020_Science,A.Li_2020_JHEA,Elhatisari:2022zrb}, as it governs a wealth of phenomena over an exceptionally broad range of scales—from terrestrial heavy-ion collisions to binary neutron star mergers and core-collapse supernovae.
Therefore, it has motivated major scientific initiatives and driven the development of key research infrastructures worldwide, including large ground-based telescopes, advanced space-based X-ray observatories, the Neutron Star Interior Composition Explorer (NICER), the LIGO/Virgo/KAGRA network of gravitational-wave detectors, and next-generation radioactive isotope beam facilities~\citep{Aasi_2015_CQG,Acernese_2015_CQG,Gendreau_2017_NA,Akutsu_2019_NA}.

Theoretically, extensive efforts have been devoted to deriving the neutron star EOS using a variety of nuclear many-body approaches, which can broadly grouped into two categories. The first category relies primarily on nuclear density functional theories (DFTs) employing effective nucleon-nucleon interactions~\citep{2020-YangJJ-ARNPS,2023-Sedrakian-PPNP}. Because the isovector sector in most DFTs is only weakly constrained during parameters optimization, their predictions for key nuclear matter quantities, such as the symmetry energy at supra-saturation densities, remain highly uncertain~\citep{Baldo_2016_PPNP,2024-Sorensen-PPNP}.

In contrast, the second category consists of \textit{ab initio} methods based on realistic forces fitted directly to few-body bound and scattering data~\citep{2021-Burgio-PPNP,2024-Machleidt-PPNP,Tong_2025_ArXiv,Tong_2025_APJ,2025-TongH-SB}. Given the extreme densities reached in neutron star interiors, a Lorentz-invariant many-body treatment is particularly advantageous, as it naturally accounts for the increasingly important relativistic kinematics and dynamics. The relativistic Brueckner–Hartree–Fock (RBHF) theory provides such a covariant framework~\citep{Shen2019PPNP}. Built upon the relativistic Bonn potentials~\citep{Machleidt1989}, RBHF theory has been successfully applied to finite nuclei, infinite nuclear matter, and neutron star calculations~\citep{Horowitz1984PLB,Brockmann_1990_PRC,Jong1998PRC,Van2005PRL,Krastev2006PRC,Katayama2013PRC,Shen_2017_PRC,Tong:2018qwx,Tong2020PRC}, offering a unified and microscopic description of strongly interacting matter across a wide density range.

Recently, a fully self-consistent RBHF framework formulated in the full Dirac space has been developed~\citep{2021-WangSB-PRC}, in which both the Lorentz structure and momentum dependence of the nucleon self-energy are determined without approximations. 
This advance resolves the long-standing inconsistencies in earlier implementations and provides a unique and accurate description of isospin effects in nuclear matter~\citep{2022-WangSB-PRCL, 2023-Wang-PhysRevC.108.L031303, 2024-WangSB-SciBull, 2024-QinPP-PRC}. 
Despite these achievements, this self-consistent RBHF calculations of neutron star EOS~\citep{2022-TongH-ApJ, 2022-WangSB-PRCL,2022-WangSB-PhysRevC.106.045804, Qu_2025-ApJ, 2025-Laskos-Patkos-PhysRevC.111.025801,2025-TongH-FASS} can currently be performed only up to baryon densities of about 0.5~fm$^{-3}$—far below the central densities of massive neutron stars. 
As a result, the EOS at higher densities remains theoretically inaccessible within the present self-consistent RBHF framework, and existing studies have relied primarily on parametric extrapolations rather than fully microscopic predictions.

Machine learning (ML) provides a powerful framework for function approximation and extrapolation, enabling one to capture complex nonlinear relationships in data that are difficult to model with traditional analytical methods. In nuclear physics, ML techniques have been successfully applied across the diversity of research topics, leading to advances that will facilitate scientific discoveries and offer deeper insights~\citep{2022-Boehnlein-RevModPhys.94.031003,2023-HeWB-ScienceChina,2024-ZhouK-PPNP}. 
In the context of nuclear matter and neutron star EOS studies, Bayesian inference techniques have been widely adopted to extract posterior distribution from terrestrial experiments, astrophysical multi-messenger observations, and microscopic nuclear theory calculations~\citep{2022-Huth-Nature,2023-HanMZ-SB,2024-Tsang-NatureAstronomy,2024-Rutherford-ApJL,Koehn_2025_PRX}. In parallel, neural networks (NNs) have been increasingly explored as powerful tools to infer EOS of dense nuclear matter from astrophysical observations of neutron stars~\citep{2018-Fujimoto-PhysRevD.98.023019,Fujimoto_2020_PRD, Traversi_2020_APJ, Morawski_2020_AA, Fujimoto_2021_JHEP, Ferreira_2021_JCAP, Krastev_2022_G, 2023-ZhouWJ-ApJ, Fujimoto_2024_PRD}. 

In previous applications of neural networks to neutron star studies, most efforts have focused on learning a direct mapping between the EOS and astrophysical observables such as the stellar mass–radius relation. As a consequence, these models provide limited access to more microscopic quantities of nuclear matter, including the symmetry energy and particle fractions. Moreover, existing studies typically incorporate the two-solar-mass constraint from the outset, which effectively forces the network to cover the entire density range during training. This setup primarily tests the interpolation capability of the network, while leaving its extrapolation performance largely unexplored. To the best of our knowledge, no work to date has directly linked neural network modeling with \textit{ab initio} nuclear matter calculations in a way that simultaneously evaluates fitting and extrapolation capabilities, and thereby yields genuine predictions for the high-density sector of the neutron star EOS.

In this work, starting from low-density datasets generated from RBHF theory in the full Dirac space, we train a neural network that predicts the binding energy per nucleon with nucleon density and isospin asymmetry as inputs. This approach enables a reliable extension of the nuclear matter EOS to high densities revelant for neutron star interiors.

This paper is organized as follows.
The theoretical framework of the RBHF theory in the full Dirac space, and the preparation of the training and validation data, together with the structure and the uncertainty estimate of the neural networks, are briefly introduced in Section \ref{SecMod}. In Section \ref{SecDisc}, the EOS, mass-radius relation, and tidal deformability of neutron star are presented. A short conclusion and outlook for future investigations are given in Section \ref{SecSUM}.

\section{Methods}\label{SecMod}

\subsection{Relativistic Brueckner-Hartree-Fock Theory in the Full Dirac Space}\label{subRBHF}

The RBHF theory provides an \textit{ab initio} and parameter-free description of nuclear matter, where the many-body correlations are treated through the effective $G$-matrix. The starting point is the single-particle Dirac equation,
\begin{equation}\label{eq:DiracEquation}
  \left[ \bm{\alpha}\cdot\bm{p}+\beta \left(M+\mathcal{U}\right) \right] u(\bm{p},s)
  = E_{\bm{p}}u(\bm{p},s),
\end{equation}
where $\bm{\alpha}$ and $\beta$ are the Dirac matrices, $M$ is the nucleon mass, $\bm{p}$ and $E_{\bm{p}}$ are the momentum and single-particle energy, and $s$ denotes the spin. The single-particle potential $\mathcal{U}$ fully incorporates the in-medium effects.

The construction of $\mathcal{U}$ relies on the $G$-matrix, which is obtained by solving the in-medium Thompson equation~\citep{Brockmann_1990_PRC},
\begin{equation}\label{eq:ThomEqu}
    \begin{split}
  G(\bm{q}',\bm{q}|\bm{P},W)
  =&\ V(\bm{q}',\bm{q}|\bm{P})
  + \int \frac{d^3k}{(2\pi)^3}
  V(\bm{q}',\bm{k}|\bm{P}) \\
    & \times 
    \frac{Q(\bm{k},\bm{P})}{W-E_{\bm{P}+\bm{k}}-E_{\bm{P}-\bm{k}} + i\epsilon}  \\
    & \times G(\bm{k},\bm{q}|\bm{P},W).
  \end{split}
\end{equation}
Here $\bm{P}$ is half of the center-of-mass momentum, and $\bm{q}$, $\bm{k}$, and $\bm{q}'$ denote the initial, intermediate, and final relative momenta respectively. The Pauli operator $Q(\bm{k},\bm{P})$ prevents scattering into occupied intermediate states.
The starting energy $W$ is taken as the sum of the single-particle energies in the initial state. This equation resums ladder diagrams to all orders and incorporates the underling one-boson-exchange interaction $V$. 



As a relativistic many-body method, RBHF theory necessitates a self-consistency solution rooted in the full Dirac space. Traditional implementations restrict the scattering equation to the subspace of positive-energy states, causing ambiguities in the scalar-vector decomposition of $\mathcal{U}$ and leading to inconsistent predictions for asymmetric nuclear matter and the symmetry energy at moderate densities.

The fully self-consistent RBHF framework in the full Dirac space developed recently overcomes these issues by treating both positive- and negative-energy states (NESs) on an equal footing~\citep{2021-WangSB-PRC, 2022-WangSB-PRCL}. This enables the construction of all matrix elements of the single-particle potential operator $\mathcal{U}$ by summing the effective two-body $G$ matrix over all nucleons within the Fermi sea in the full Dirac space. This full Dirac space construction allows for the exact extraction of the Lorentz structure of $\mathcal{U}$, which is crucial for solving the Dirac equation in the next iteration. Consequently, inclusion of the NESs has not only clarified the long-standing controversy about the isospin dependence of the effective mass in neutron-rich matter~\citep{2022-WangSB-PRCL}, but also yielded a fully consistent determination of the single-particle spectrum~\citep{2023-Wang-PhysRevC.108.L031303} and the in-medium interaction~\citep{2025-WangTY-Phys.Rev.C}.

With the converged Dirac spinor and $G$-matrix, the binding energy per nucleon of nuclear matter is calculated by
\begin{equation}
  \begin{split}
  E/A
  =&\ \frac{1}{\rho} \int_{\bm{p}} \langle
    \bar{u}(\bm{p})| \bm{\gamma}\cdot\bm{p} + M |u(\bm{p})\rangle - M \\
    &\ + \frac{1}{2\rho} \int_{\bm{p}} \int_{\bm{p}'}
     \langle \bar{u}(\bm{p}) \bar{u}(\bm{p}') |\bar{G}(W)| u(\bm{p}) u(\bm{p}') \rangle.
  \end{split}
\end{equation}
Here, \textbf{}the density $\rho$ is related to the Fermi momentum $k_\text{F}$ through $\rho=2k^3_\text{F}/3\pi^2$, $\int_{\bm{p}} \equiv \int^{k_\text{F}}_0 d^3p/(2\pi)^3$, $W=E_{\bm{p}}+E_{\bm{p}'}$, and the spin as well as isospin indices are suppressed. More theoretical and numerical details are provided in~\cite{2021-WangSB-PRC, 2022-WangSB-PRCL, 2022-TongH-ApJ}.

In the vicinity of the nuclear saturation density, the binding energy per nucleon of asymmetric nuclear matter (ANM) can be generally expressed as a power series in the asymmetry parameter $\alpha= (\rho_{n}-\rho_{p})/\rho $, with $\rho_{n}, \rho_{p}$, and $\rho = \rho_{n} + \rho_{p}$ being the neutron, proton, and nucleon density, respectively  \citep{Bombaci_1991_PRC}:
\begin{equation}\label{equa1}
E/A(\rho,\alpha) = E_{\rm SNM}(\rho) + E_{\rm sym}(\rho)\alpha^{2} + \mathcal{O}(\alpha^4).
\end{equation}
Here, $E_{\rm SNM}(\rho) = E/A(\rho, 0)$ is the binding energy per nucleon of symmetric nuclear matter (SNM), and $E_{\rm sym}(\rho)$ is the density-dependent symmetry energy.
Presently, the EOS of cold nuclear matter under extreme conditions of density and isospin asymmetry still remains rather uncertain and theoretically controversial, particularly at supra-saturation densities and with extreme neutron abundance, mainly due to the poorly known high-density behavior of the nuclear symmetry energy $E_{\rm sym}(\rho)$ \citep{Li_2014_EPJA,Li_2021_U}. The RBHF theory formulated in the full Dirac space thus provides a unique and ambiguity-free microscopic description of nuclear matter up to densities $\rho\approx 0.5\ \text{fm}^{-3}$. This makes it an ideal generator of high-fidelity, first-principle data, enabling our ML model to extrapolate the EOS into the high-density region relevant for neutron star interiors.

\subsection{Training and Validation Data}\label{subdata}

The neutral networks developed in this work are trained on high-fidelity datasets generated from the fully self-consistent RBHF calculations in the full Dirac space. These \textit{ab initio} calculations are computationally demanding and can only be stably converged up to densities $\rho\approx 0.52-0.57\ \text{fm}^{-3}$, depending on the isospin asymmetry~\citep{2022-WangSB-PRCL}. Owing to its larger Fermi momentum, pure neutron matter reaches its maximum attainable density at a lower value than SNM. As the computational cost increases sharply with the increase in density, the available dataset is sparse in density and uneven in $\alpha$, while remaining extremely reliable in the converged region. This motivates a training strategy explicitly designed to preserve the predictive power of the RBHF inputs while enabling controlled extrapolation. 

The data are sampled on a fixed density grid,
\begin{equation}
    \rho_i = 0.08 + 0.01 i\ \text{fm}^{-3},\quad i = 0, 1, \cdots, i_\text{max},
\end{equation}
extending slightly above $0.5\ \text{fm}^{-3}$, and on eleven asymmetry parameters
\begin{equation}
    \alpha_j = 0.1j,\quad j = 0,1,\cdots, 10.
\end{equation}
For every sampled pair $(\rho_i, \alpha_j)$, the corresponding binding energy per nucleon $E/A(\rho_i, \alpha_j)$ is obtained. In total, around 530 converged points are available.

Since the ultimate goal is to extrapolate the EOS into the high-density regime relevant for neutron star interiors, the dataset must allows us to rigorously assess the model's robustness specifically under high densities and large isospin asymmetry. To this end, the partition of training and validation sets is constructed as follows:
\begin{equation}
    \begin{split}
        \mathcal{D}_{\text{train}} =&\ \{E/A(\rho, \alpha) | \alpha\leq 0.7,\ \rho \leq 0.5\ \text{fm}^{-3}\} \\
            &\quad \cup \ \{E/A(\rho, \alpha) | \alpha\geq 0.8,\ \rho \leq 0.45\ \text{fm}^{-3}\}, \\
        \mathcal{D}_{\text{validation}} =&\ \text{remaining data with larger $\rho$ at each $\alpha$}.
    \end{split}
\end{equation}
This partitioning yields 458 and 72 data points for the training and validation sets, respectively. Both density $\rho$ and isospin asymmetry $\alpha$ are rescaled to the interval $[-1,1]$ for numerical stability, and the binding energy per nucleon is normalized by a constant scale. No smoothing, fitting, or interpolation is applied and the neural network learns directly from the raw RBHF calculations. This preserves the \textit{ab initio} nature of the input and ensures that all features learned by the model originate from the underlying microscopic theory.

\subsection{Neural Network Architecture}



\begin{figure}[htbp]
\centering
\includegraphics[width=8.5cm]{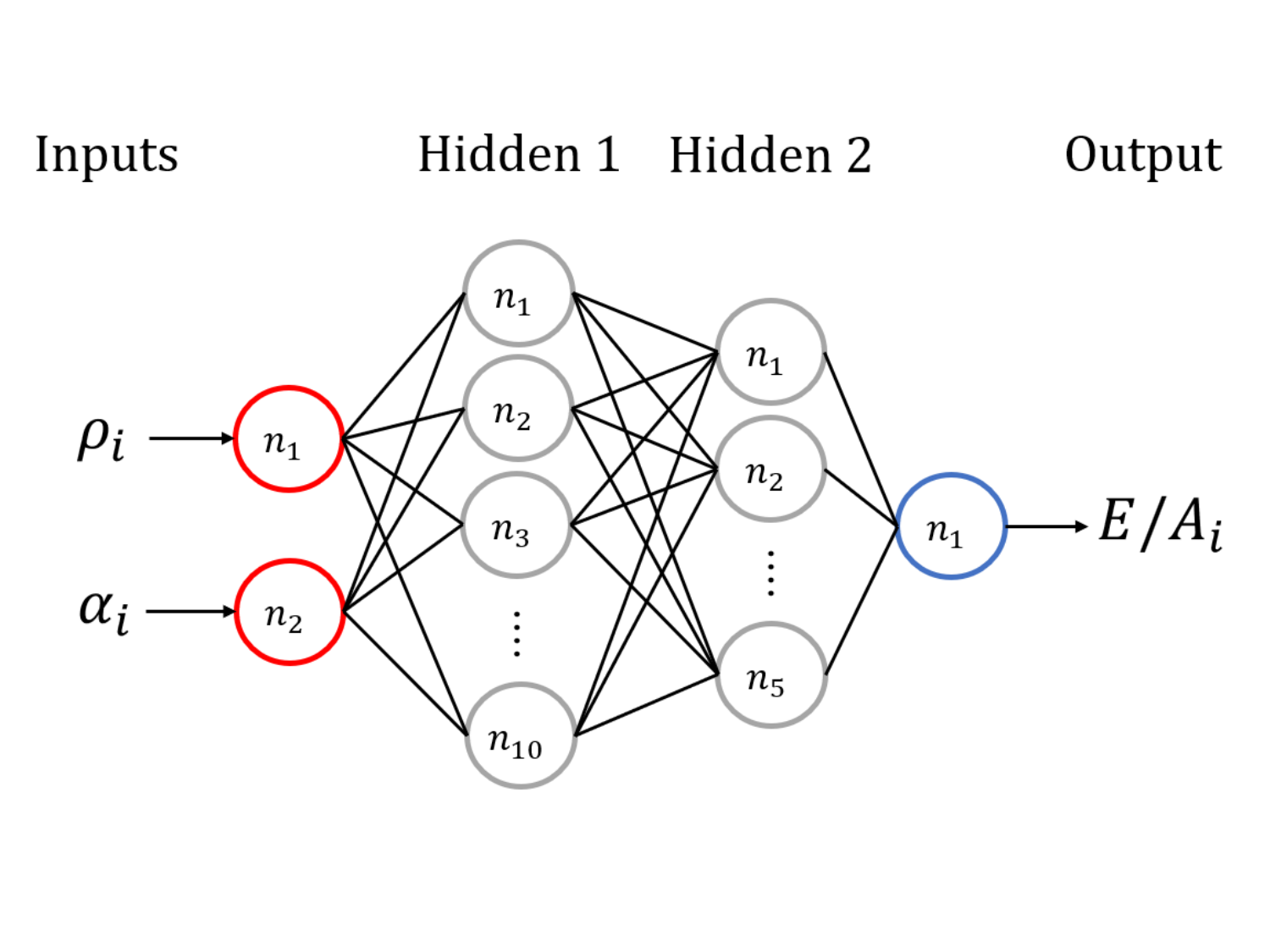}
\caption{Schematic illustration of the feedforward neural network.}
\label{Fig1NN}
\end{figure}

To emulate the density and isospin dependence of the RBHF EOS, we construct a fully connected neural network that maps the two physical input variables, the nucleon density $\rho$ and the isospin asymmetry $\alpha$, to a single output, the binding energy per nucleon $E/A(\rho, \alpha)$. The architecture of our neural network is illustrated in Figure~\ref{Fig1NN}, consisting of an input layer with two neurons, two hidden layers with 10 and 5 neurons respectively, and a single-neuron output layer. All networks in the ensemble share this structure, differing only in the realization of randomly initialized weights and biases. 

Unlike many applications of ML where the choice of activation function impacts only performance, in our case the activation function plays a central physical role: it governs whether the learned EOS can be meaningfully extrapolated to densities beyond the RBHF convergence limit. The activation function must therefore satisfy two criteria: (1) The network must accurately learn the smooth dependence of $E/A(\rho, \alpha)$ encoded in the RBHF data across the sampled range of densities and asymmetries. (2) When extended to higher densities relevant for neutron star interiors, the pressure $P(\rho)$ inferred from the neural work must obey: (i) Thermodynamic stability $dP/d\rho>0$; (ii) smoothness since our model contains neither phase transitions nor additional degrees of freedom.

These stringent physical conditions significantly constrain the choice of viable activation functions. We therefore conducted a systematic evaluation of commonly used activation functions including ReLu, Sigmoid, and Tanh. In addition, we also test the Swish activation function~\citep{Prajit_2018_ArXiv}, 
\begin{equation}
  \mathrm{Swish}(x) = x \cdot \frac{1}{1 + e^{-\beta x}},
\end{equation}
which is not commonly used in EOS-related neural network studies. We set $\beta = 1$ in this work~\citep{Prajit_2018_ArXiv}. Each activation function was evaluated based on its accuracy on the RBHF training and validation data at low and moderate densities and the physical plausibility of the extrapolated EOS at high densities. The computations are performed in Python, employing Keras~\citep{Chollet_2015} with TensorFlow~\citep{Martín_2016_arXiv} as the backend. For optimization, the Adam algorithm~\citep{Diederik_2017_ArXiv} is adopted with a batch size of 32, and the mean squared error is adopted as the loss function.



\begin{figure}[htbp]
\centering
\includegraphics[width=8.5cm]{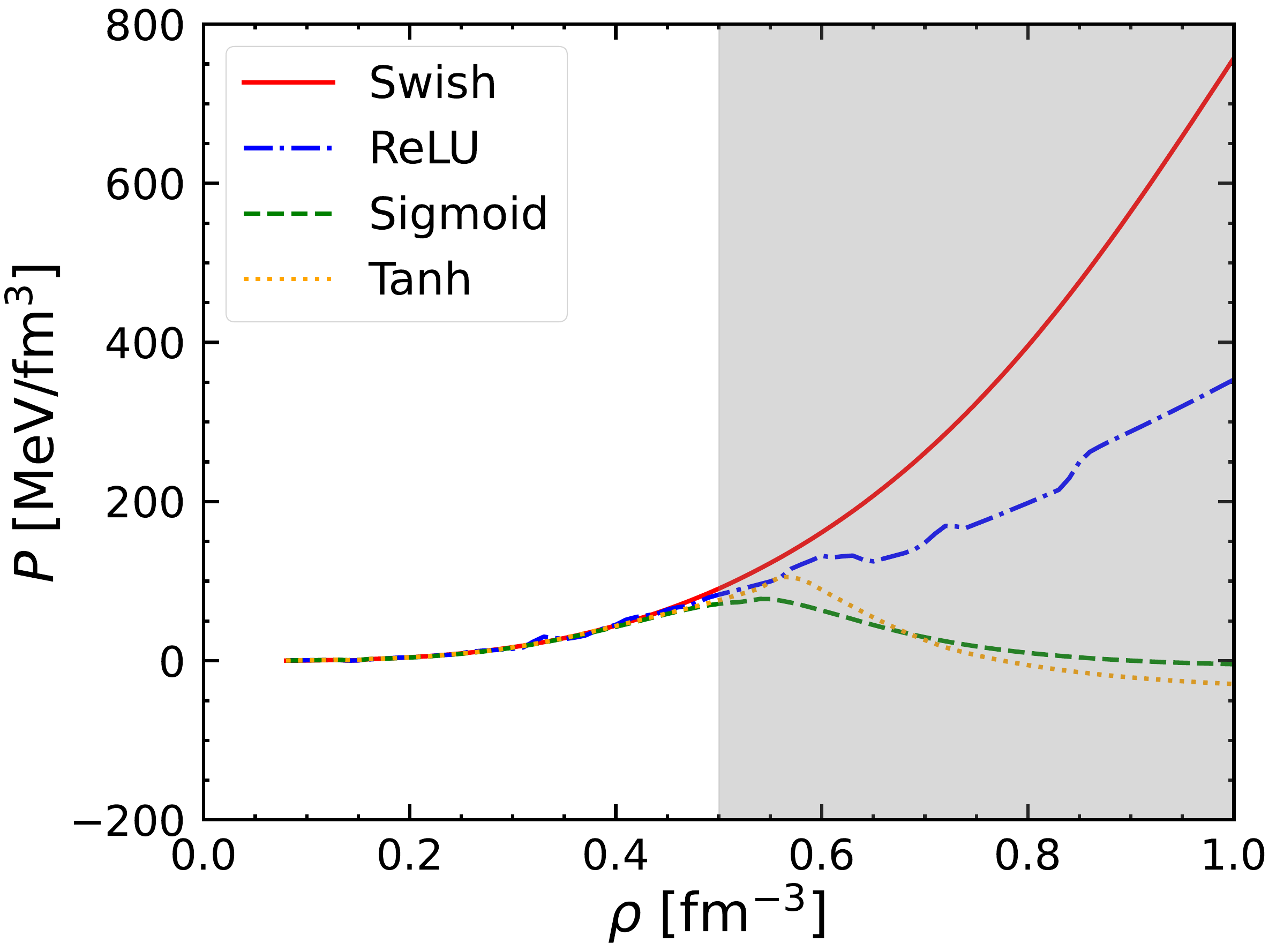}
\caption{Pressure of neutron star matter as a function of nucleon number density obtained using four activation functions: Swish (red solid line), ReLu (blue dash-dotted line), Sigmoid (green dashed line) and  Tanh (orange dotted line). The white area indicates the part of the dataset used for training, whereas the gray-shaded area represents the extrapolation region where the model predictions are evaluated.}
\label{Fig2ACT}
\end{figure}

Figure~\ref{Fig2ACT} depicts the pressure of neutron star matter as a function of nucleon density calculated with $E/A(\rho, \alpha)$ obtained with four activation functions. Our tests reveal that the Swish activation function offers distinctly superior extrapolation behavior. ReLU networks tend to produce non-smooth pressures and even thermodynamic instabilities, and Sigmoid/Tanh saturate too rapidly to represent the high-density rise of the EOS. In comparison, Swish network captures the continuous growth of the nuclear matter EOS at supra-saturation densities, which potentially corresponds to the repulsive behavior of nucleon-nucleon interactions in the high-density region. As shown in Figure~\ref{Fig2ACT}, only Swish yields an EOS that remains smooth and thermodynamically stable throughout the extrapolation region. Consequently, all neural networks selected for the final ensemble employ Swish in all hidden layers. 

The superior performance of the Swish activation function is attributed to its ability to mitigate critical flaws in both ReLU and saturation functions like Sigmoid/Tanh. Specifically, it circumvents the instability and neuron "death" associated with ReLU during extrapolation and, unlike Sigmoid/Tanh, does not suffer from complete gradient vanishing in unseen data regions~\citep{Apicella_2021_NN,Goodfellow_2016_MIT}. Consequently, Swish exhibits substantially more stable and robust predictive behavior. Furthermore, its global smoothness and full differentiability make it a more effective alternative than ReLU, Sigmoid, or Tanh for tasks that require robust generalization and accurate predictions beyond the training data distribution.

\subsection{Uncertainty Estimation}\label{subuncer}

\begin{figure}[htbp]
\centering
\includegraphics[width=8.5cm]{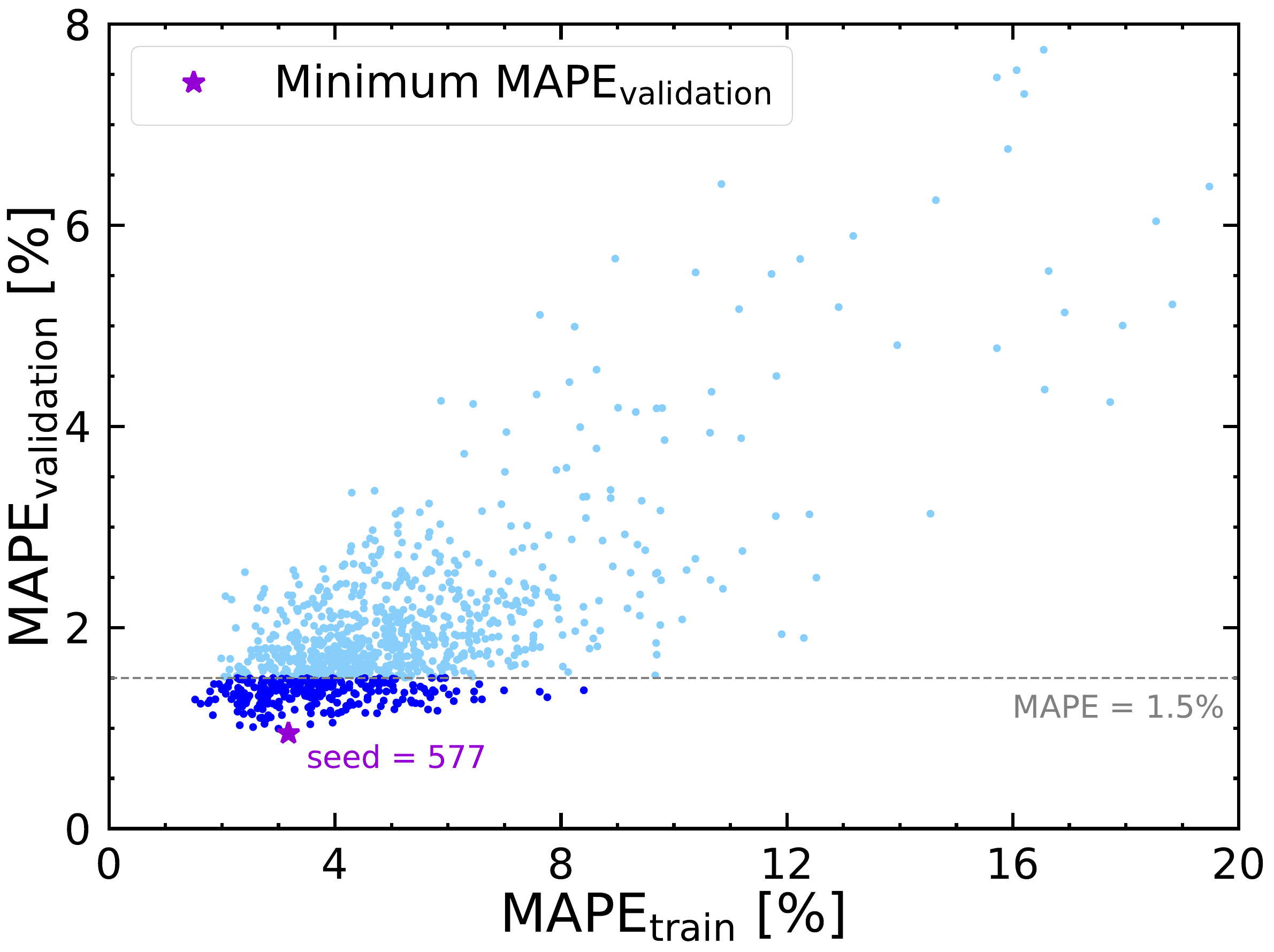}
\caption{Distribution and selection of 1000 neural networks based on MAPE. The dashed line indicates the 1.5\% validation MAPE threshold. Minimum validation MAPE (0.95$\%$) achieved by the neural network with random seed 577 is marked with a purple star.}
\label{Fig3MAPE}
\end{figure}

The stochastic initialization of weights and biases during neural network training inevitably introduces a degree of randomness into the model, leading to slightly different mappings across different training runs. To quantify the resulting model uncertainty and ensure reproducibility, we construct an ensemble of 1000 independently trained neural networks, each initialized with a distinct but fixed random seed. For every model in the ensemble, we evaluate the agreement with the RBHF data using the mean absolute percentage error (MAPE),
\begin{equation}
    \text{MAPE} = \frac{1}{N} \sum_{i=1}^N \left| \frac{(E/A)_{\text{NN},i}-(E/A)_{\text{RBHF},i}}{(E/A)_{\text{RBHF},i}}\right|,
\end{equation}
where the sum runs over the $N$ data points. This metric provides a dimensionless and robust measure of the model's extrapolation accuracy, independent of the absolute scale of the binding energy.

The distribution of MAPE values for the training and validation sets across the ensemble is shown in Figure \ref{Fig3MAPE}. A clear positive correlation is observed. Neural networks those fit the training data more accurately also exhibit smaller validation errors, indicating that overfitting is negligible and the learned functional dependence is physically robust. To isolate models with reliable extrapolation behavior, we impose a threshold $\mathrm{MAPE_{validation}}\leq 1.5\%$. In total 264 neural networks satisfy this criterion, and only these are retained for all subsequent predictions of the symmetry energy, pressure, and neutron star observables. Notably, the neural network corresponding to random seed 577 yields the minimum validation MAPE (0.95$\%$) and is indicated by a purple star in Figure \ref{Fig3MAPE}. The spread among this selected ensemble provides a direct statistical estimate of the theoretical uncertainty associated with the ML extrapolation.

\section{Results and discussions}\label{SecDisc}

The exceptional extrapolative capability of our neural network model, demonstrated through its predictions of the binding energy per nucleon across a wide density range, paves the way for a unified calculation of key nuclear and astrophysical properties. This enables us to derive the nuclear matter symmetry energy from low to high densities, the neutron star matter pressure, and critical neutron star observables—encompassing the mass-radius relation and tidal deformability.

\begin{figure*}[htbp]
\centering
\includegraphics[width=0.9\textwidth]{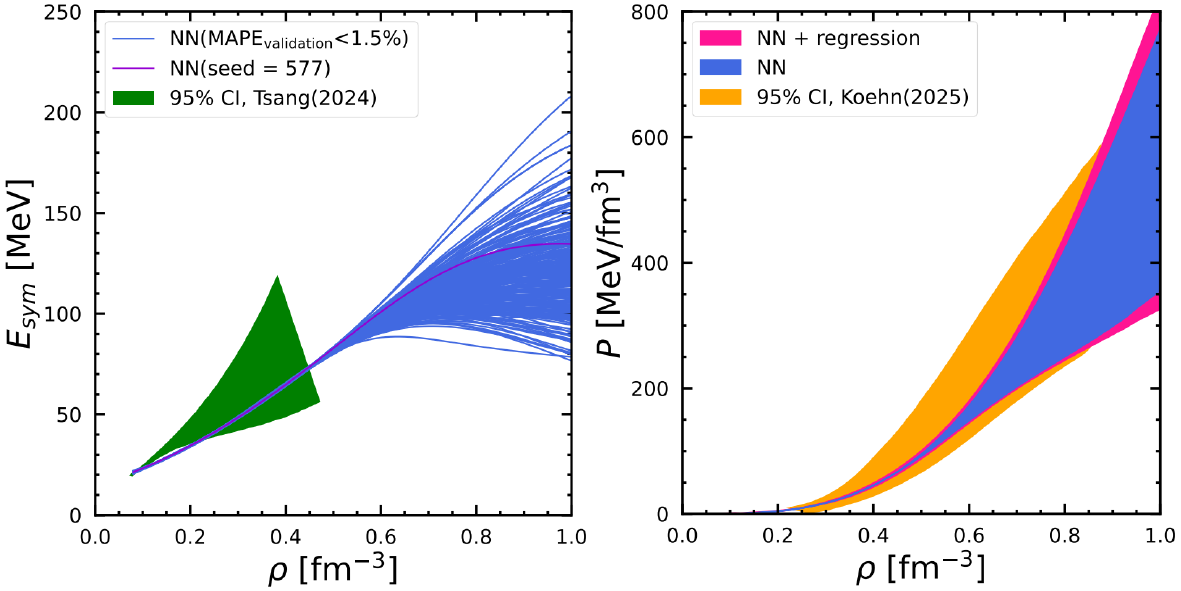} \\
\caption{Left Panel: Nuclear matter symmetry energy $E_\mathrm{{sym}}$ as a function of density $\rho$. The blue solid lines denote the neural networks satisfying $\mathrm{MAPE_{validation}} \le 1.5 \%$, while the purple solid line denotes the one with the minimum $\mathrm{MAPE_{validation}}$. The  green shaded regions represent $95\%$ confidence interval (CI) of the posterior distributions from~\textcite{2024-Tsang-NatureAstronomy}.
Right Panel: Neutron star matter pressure $P$ as a function of density $\rho$. The blue shaded region corresponds to the uncertainty obtained from the neural networks with $\mathrm{MAPE_{validation}} \le 1.5\%$, while the red shaded regions correspond to the $95 \%$ CI of $E_{\rm sym}$, derived from the simple linear regression in Equation~\eqref{equa1}. The orange shaded regions correspond to the $95\%$ CI of the posterior distribution from~\textcite{Koehn_2025_PRX}.} 
\label{Fig4EP}
\end{figure*}

It should be emphasized that the neural networks are trained to predict the binding energy per nucleon $E/A(\rho,\alpha)$ for arbitrary density and isospin asymmetry. To obtain the symmetry energy at a fixed density, we perform a linear regression fitting the machine-learning predictions to the quadratic expansion in Equation~\eqref{equa1}. This regression approach makes use of the full information provided by the neural network EOS with eleven asymmetry values $\alpha = 0, 0.1, \cdots, 1.0$, thus avoids relying solely on the two extreme points with the commonly used approximation $E_\text{sym} (\rho) \approx E/A (\rho, \alpha = 1) - E/A (\rho, \alpha = 0)$, which becomes worse for higher densities \citep{Wang_2020_APJ}. In addition, the residual contribution from high-order terms in Equation~\eqref{equa1} is effectively incorporated into the statistical uncertainty of the fitted coefficient. 
As a result, the extracted $E_\text{sym}(\rho)$ is more robust, less biased, and more faithful to the underlying functional dependence learned by the neural networks.

The resulting symmetry energy curves for the selected ensemble (blue dots in Figure~\ref{Fig3MAPE}) are shown as blue lines in Figure~\ref{Fig4EP}. Up to $\rho\approx 0.5\ \text{fm}^{-3}$ where RBHF calculations provide direct microscopic input, the predictions from different neural networks are nearly indistinguishable, indicating that the regression reliably captures the RBHF trend. Our calculations are also comparable to the results of Bayesian inference that incorporate astronomical observations and nuclear experimental constrains \citep{2024-Tsang-NatureAstronomy}. Beyond $0.5~\mathrm{fm^{-3}}$, however, the spread among the models grows rapidly, reflecting the well-known difficulty of constraining the symmetry energy at supra-saturation densities. Up to $0.8~\mathrm{fm^{-3}}$, although a small number of neural networks predict a saturation or even a mild decline of at very high densities, the majority exhibit a steadily increasing trend. Such behavior is consistent with expectations from microscopic Brueckner–Hartree–Fock calculations based on realistic nuclear interactions as well as from relativistic mean-field models employing phenomenological forces \citep{2008-LiBA-Phys.Rep.}. The growing dispersion in the extrapolated region therefore quantifies the current theoretical uncertainty. Specifically, the symmetry energy at empirical and twice saturation density are $E_\text{sym}(\rho_0) = 29.4 \pm  0.3 $\ MeV, $E_\text{sym}(2\rho_0) = 51.6 \pm 3.4$\ MeV, respectively. Both are in agreement with experimental constraints $E_\text{sym}(\rho_0) = 31.7 \pm  3.2 $\ MeV \citep{Oertel_2017_RMP} and $E_\text{sym}(2\rho_0) = 55.6 \pm 4.8$\ MeV from ASY-EOS \citep{Russotto_2016_PRC}. Extending to five times saturation density, our neural networks studies predict $E_\text{sym}(5\rho_0) = 136.0 \pm 52.8$\ MeV, which is left to be compared with other theoretical or experimental studies.

Based on the density dependent symmetry energy obtained above, we next construct the pressure of cold $\beta$-equilibrated neutron star matter. For each neural-network EOS, particle fractions are determined by solving simultaneously the conditions of chemical equilibrium and charge neutrality. With the composition fixed, the total energy density and pressure are computed in the standard way. 

\begin{figure*}[htbp]
\centering
\includegraphics[width=0.9\textwidth]{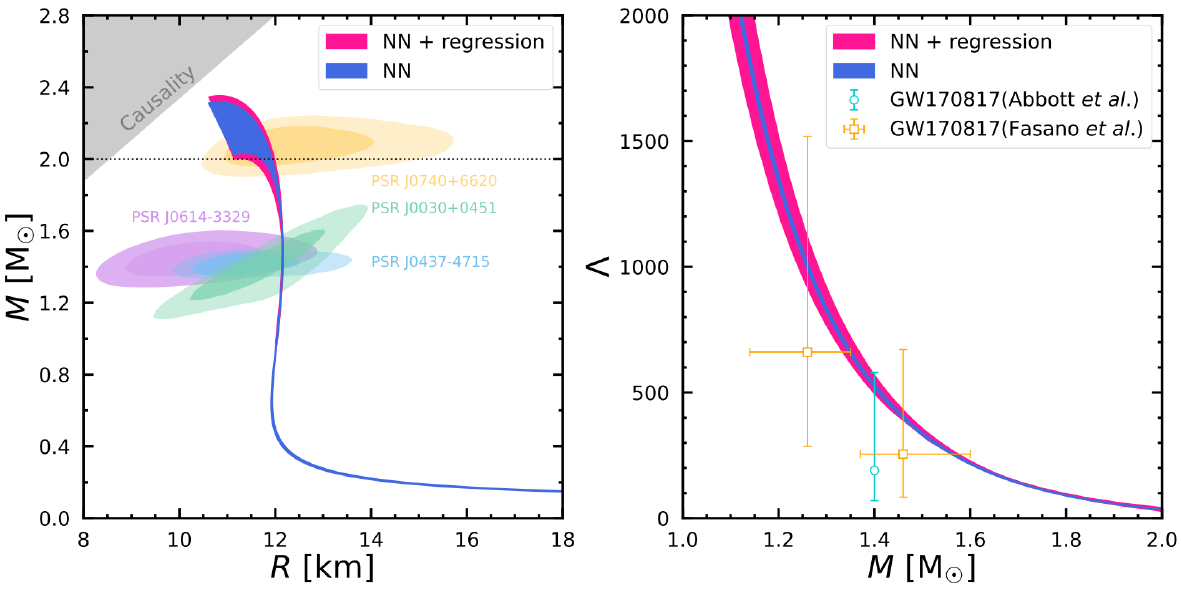} \\
\caption{Left Panel: Neutron star mass-radius relation. 
The blue and red bands represent the uncertainties from the neural networks alone and from the combined NN and linear regression, respectively.
The horizontal dotted line marks 2$M_{\odot}$. The inner and outer shaded contours indicate the mass-radius constraints from NICER's analysis of PSR J0030+0451 \citep{Vinciguerra_2024_APJ}, PSR J0740+6620 \citep{Salmi_2024_APJ}, PSR J0437-4715 \citep{Choudhury_2024_APJL} and PSR J0614-3329 \citep{Lucien_2025_ArXiv}.
Right Panel: Neutron star tidal deformability $\Lambda$ as a function of mass. Our results (blue and red regions) are compared with the tidal deformabilities for the two neutron stars in the merger event GW170817 as reported in \cite{Margherita_2019_PRL} (open squares), as well as the value of $\Lambda_{1.4M_{\odot}}$ extracted from GW170817 \citep{Abbott_2018_PRL} (open circle).}
\label{Fig5MR}
\end{figure*}

The resulting pressure bands are shown in the right panel of Figure~\ref{Fig4EP}. Two types of uncertainties are displayed: the blue band reflects the spread among the selected neural networks alone, while the red band includes, in addition, the statistical uncertainty associated with the linear regression used to extract the symmetry energy at each density. It is evident that the neural network variability dominates the total uncertainty, while the contribution from the regression analysis is comparatively small. Qualitatively, all selected models predict a monotonically increasing pressure with density and show no sign of saturation even at the highest densities considered. This behavior is consistent with the thermodynamic-stability criterion imposed when selecting the activation function: the extrapolated EOS must be stable and remain smooth in the absence of phase transitions or additional degrees of freedom. Below $0.8~\mathrm{fm^{-3}}$, our results fall well within the Bayesian posterior distributions (specifically, Set A) of \textcite{Koehn_2025_PRX}, which were derived by incorporating constraints from microscopic theoretical calculations and multi-messenger astronomy observations with minimal model dependence. Quantitatively, our pressure at five times the empirical saturation density is found to be $P(5\rho_0)= 346.3 \pm 97.4 ~\mathrm{MeV/fm^{3}}$.

With the EOS of $\beta$-equilibrated neutron star matter established in the previous section supplemented with BPS-BBP model~\citep{1971-BPS-ApJ.170.299B,1971-BBP-Nucl.Phys.A175.225} for the crust, we compute the stellar structure by solving the Tolman–Oppenheimer–Volkoff (TOV) equations for hydrostatic equilibrium.
In parallel, the dimensionless tidal deformability $\Lambda$ is obtained by integrating the standard first-order differential equation describing quadrupolar metric perturbations, coupled to the TOV background. Together, these solutions yield the mass–radius relation and tidal deformabilities corresponding to each neural network EOS, as shown in Figure~\ref{Fig5MR}.

For the mass–radius relation (left panel), the maximum mass supported by the ensemble is $2.18 \pm 0.18 M_{\odot}$, including both the neural network and linear regression uncertainties. Although the uncertainty band is relatively broad, it is noteworthy that all neural networks predict a maximum mass above $2.0 M_{\odot}$, fully consistent with the heavy pulsars observed in radio timing measurements \citep{Salmi_2024_APJ,Demorest_2010_Nature,Fonseca_2016_APJ,Arzoumanian_2018_APJSS}. This agreement confirms that the neural networks remain stable and physically reliable during extrapolation to the high-density regime where no RBHF data are available. 

At canonical mass, the radius of a $1.4 M_{\odot}$ neutron star is predicted to lie within a very narrow range around $R_{1.4M_{\odot}}=12 ~\text{km}$, reflecting the fact that its central density $\approx 0.46\ \text{fm}^{-3}$~\citep{2022-TongH-ApJ}, does not exceed the region where the selected neural networks show appreciable divergence. In contrast, the radius at $2.0 M_{\odot}$ is found to be $R_{2.0M_{\odot}} = 11.8 \pm 0.3 ~\text{km}$, which spans a noticeably wider interval, since these stars probe the higher-density regime above $5-6\rho_0$ where extrapolation uncertainties naturally increase. Importantly, the mass–radius curves remain fully compatible with the NICER measurements of PSR~J0030+0451 \citep{Vinciguerra_2024_APJ}, PSR~J0740+6620 \citep{Salmi_2024_APJ}, PSR~J0437$-$4715 \citep{Choudhury_2024_APJL}, and PSR~J0614$-$3329 \citep{Lucien_2025_ArXiv}, as illustrated by the observational contours in Figure~\ref{Fig5MR}. Therefore, the RBHF-trained machine learning EOS provides a realistic description of dense matter, as it leverages the high fidelity of RBHF at low and intermediate densities and maintains reliable extrapolation to higher densities through the use of physically motivated activation function.


The right panel of Figure~\ref{Fig5MR} displays the corresponding tidal deformabilities. The tidal deformability $\Lambda$ quantifies the ease with which a neutron star develops a quadrupole moment in response to an external tidal field, and is therefore a key observable in multi-messenger astronomy \citep{Abbott_2017_PRL}. At present, GW170817 remains the only binary neutron star merger from which $\Lambda$ constraints have been obtained. Since gravitational-wave observations determine primarily the chirp mass, the inferred constraints appear as a joint allowed region for the two stellar masses and their tidal deformabilities. Our neural network predictions fall squarely within this allowed region, demonstrating consistency with current observational limits.

For $1.4 M_{\odot}$ neutron star, we obtain $\Lambda_{1.4M_\odot}=532\pm34$, lying toward the upper edge of the GW170817 constraint $\Lambda_{1.4M_\odot}=190^{+390}_{-120}$ \citep{Abbott_2018_PRL}. This placement indicates that the gravitational-wave analysis tends to favor a comparatively soft EOS at low densities, producing smaller stellar radii and hence smaller tidal deformabilities. When combined with the observational requirement that the EOS must also support neutron stars of at least $2.0M_\odot$, a picture emerges that EOS must undergo a relatively rapid stiffening between the densities characteristic of canonical mass stars and those associated with the most massive neutron stars. Such a transition from low-density softness to high-density hardness is one of the most intriguing and currently unresolved features in neutron star matter research.
 
To better represent the regime where EOS uncertainties become most significant, we also quote the deformability of a $2.0M_\odot$ neutron star, for which the central density reaches into the extrapolated domain: $\Lambda_{2.0M_\odot} = 31\pm10$. Future gravitational-wave detections of heavier binary systems will be essential for constraining this high-density region and clarifying whether the rapid stiffening suggested by current multi-messenger data is indeed a robust feature of dense matter.

\section{Conclusions}\label{SecSUM}

We present a neural network construction of the equation of state (EOS) based on relativistic \textit{ab initio} input. By training an ensemble of networks on relativistic Brueckner-Hartree-Fock (RBHF) data and imposing physical constraints, we obtain an EOS that is consistent with microscopic calculations at low and intermediate density and provides a stable, causal extension to the high-density region. The success of this extrapolation hinges on the use of the Swish activation function.

The density dependence of the symmetry energy, extracted through a regression analysis of the machine-learned EOS, exhibits excellent consistency among neural networks up to $\rho=0.5\ \text{fm}^{-3}$. Beyond this region, the model spread increases, reflecting the intrinsic uncertainty of the high-density symmetry energy. The pressure of cold $\beta$-equilibrated matter, obtained after solving chemical-equilibrium and charge-neutrality conditions, shows that the neural network variation dominates the total uncertainty.

When applied to neutron star structure, the extended EOS supports maximum masses above $2.0M_\odot$ and yields mass–radius relations fully compatible with NICER observations. The predicted tidal deformabilities also fall within the broad range inferred from GW170817, with $\Lambda_{1.4M_\odot}$ lying near the upper envelope of the allowed interval, consistent with a relatively soft low-density EOS that must stiffen rapidly to support heavy neutron stars. Such a soft-to-stiff transition is a notable feature of current multi-messenger-constrained EOSs.

Our results demonstrate that combining relativistic \textit{ab initio} input with physically constrained machine-learning extrapolation provides a robust and predictive framework for dense-matter studies. Moreover, this methodology can be readily extended to incorporate other \textit{ab initio} calculations based on chiral effective field theory. This extension would facilitate reliable extrapolations across density regions and contribute to a unified microscopic understanding of the neutron star EOS.



\section{Acknowledgments}\label{Acknow}

HT acknowledges funding by the European Research Council (ERC) under the European Union's Horizon 2020 research and innovation programme (AdG EXOTIC, Grant agreement No. 101018170) and by the MKW NRW under the funding code NW21-024-A.
This work was partly supported by the National Natural Science Foundation of China under Grant Nos. 12205030, 12575130, 12265012, Chongqing Natural Science Foundation under Grant No. CSTB2025NSCQ-GPX0742, and Project of Guizhou Provincial Department of Science and Technology under Grant No.CXTD[2025]030.
Part of this work was achieved by using the supercomputer OCTOPUS at the Cybermedia Center, Osaka University under the support of Research Center for Nuclear Physics of Osaka University.

\bibliography{RBHF}

@article{Tong:2018qwx,
    author = "Tong, Hui and Ren, Xiu-Lei and Ring, Peter and Shen, Shi-Hang and Wang, Si-Bo and Meng, Jie",
    title = "{Relativistic Brueckner-Hartree-Fock theory in nuclear matter without the average momentum approximation}",
    eprint = "1808.09138",
    archivePrefix = "arXiv",
    primaryClass = "nucl-th",
    doi = "10.1103/PhysRevC.98.054302",
    journal = "Phys. Rev. C",
    volume = "98",
    number = "5",
    pages = "054302",
    year = "2018"
}

@article{Elhatisari:2022zrb,
    author = "Elhatisari, Serdar and others",
    title = "{Wavefunction matching for solving quantum many-body problems}",
    eprint = "2210.17488",
    archivePrefix = "arXiv",
    primaryClass = "nucl-th",
    doi = "10.1038/s41586-024-07422-z",
    journal = "Nature",
    volume = "630",
    number = "8015",
    pages = "59--63",
    year = "2024"
}

@MISC{Martín_2016_arXiv,
  author = {Martín Abadi and Paul Barham and Jianmin Chen and Zhifeng Chen and
	Andy Davis and Jeffrey Dean and Matthieu Devin and Sanjay Ghemawat
	and Geoffrey Irving and Michael Isard and Manjunath Kudlur and Josh
	Levenberg and Rajat Monga and Sherry Moore and Derek G. Murray and
	Benoit Steiner and Paul Tucker and Vijay Vasudevan and Pete Warden
	and Martin Wicke and Yuan Yu and Xiaoqiang Zheng},
  title = {TensorFlow: A system for large-scale machine learning},
  year = {2016},
  archiveprefix = {arXiv},
  eprint = {1605.08695},
  primaryclass = {cs.DC},
  url = {https://arxiv.org/abs/1605.08695}
}

@article{2022-Huth-Nature,
  author = {Huth, Sabrina and Pang, Peter T. H. and Tews, Ingo and Dietrich, Tim and Le Fèvre, Arnaud and Schwenk, Achim and Trautmann, Wolfgang and Agarwal, Kshitij and Bulla, Mattia and Coughlin, Michael W. and Van Den Broeck, Chris},
  title = {Constraining Neutron-Star Matter with Microscopic and Macroscopic Collisions},
  journal = {Nature},
  volume = {606},
  number = {7913},
  pages = {276--280},
  year = {2022},
  month = {jun},
  date = {2022/06/01},
  issn = {1476-4687},
  doi = {10.1038/s41586-022-04750-w},
  url = {https://doi.org/10.1038/s41586-022-04750-w}
}

@article{2024-Tsang-NatureAstronomy,
  author = {Tsang, Chun Yuen and Tsang, ManYee Betty and Lynch, William G. and Kumar, Rohit and Horowitz, Charles J.},
  title = {Determination of the Equation of State from Nuclear Experiments and Neutron Star Observations},
  journal = {Nature Astronomy},
  volume = {8},
  number = {3},
  pages = {328--336},
  year = {2024},
  month = {mar},
  date = {2024/03/01},
  issn = {2397-3366},
  doi = {10.1038/s41550-023-02161-z},
  url = {https://doi.org/10.1038/s41550-023-02161-z}
}

@article{2023-HanMZ-SB,
title = {Plausible presence of new state in neutron stars with masses above 0.98MTOV},
journal = {Science Bulletin},
volume = {68},
number = {9},
pages = {913-919},
year = {2023},
issn = {2095-9273},
doi = {https://doi.org/10.1016/j.scib.2023.04.007},
url = {https://www.sciencedirect.com/science/article/pii/S2095927323002475},
author = {Ming-Zhe Han and Yong-Jia Huang and Shao-Peng Tang and Yi-Zhong Fan}
}

@ARTICLE{Abbott_2017_PRL,
  author = {Abbott, B. P. and Abbott, R. and Abbott, T. D. and Acernese, F. and
	Ackley, K. and Adams, C. and Adams, T. and Addesso, P. and Adhikari,
	R. X. and Adya, V. B. and Affeldt, C. and Afrough, M. and Agarwal,
	B. and Agathos, M. and Agatsuma, K. and Aggarwal, N. and Aguiar,
	O. D. and Aiello, L. and Ain, A. and Ajith, P. and Allen, B. and
	Allen, G. and Allocca, A. and Altin, P. A. and Amato, A. and Ananyeva,
	A. and Anderson, S. B. and Anderson, W. G. and Angelova, S. V. and
	Antier, S. and Appert, S. and Arai, K. and Araya, M. C. and Areeda,
	J. S. and Arnaud, N. and Arun, K. G. and Ascenzi, S. and Ashton,
	G. and Ast, M. and Aston, S. M. and Astone, P. and Atallah, D. V.
	and Aufmuth, P. and Aulbert, C. and AultONeal, K. and Austin, C.
	and Avila-Alvarez, A. and Babak, S. and Bacon, P. and Bader, M. K. M.
	and Bae, S. and Bailes, M. and Baker, P. T. and Baldaccini, F. and
	Ballardin, G. and Ballmer, S. W. and Banagiri, S. and Barayoga, J. C.
	and Barclay, S. E. and Barish, B. C. and Barker, D. and Barkett,
	K. and Barone, F. and Barr, B. and Barsotti, L. and Barsuglia, M.
	and Barta, D. and Barthelmy, S. D. and Bartlett, J. and Bartos, I.
	and Bassiri, R. and Basti, A. and Batch, J. C. and Bawaj, M. and
	Bayley, J. C. and Bazzan, M. and Bécsy, B. and Beer, C. and Bejger,
	M. and Belahcene, I. and Bell, A. S. and Berger, B. K. and Bergmann,
	G. and Bernuzzi, S. and Bero, J. J. and Berry, C. P. L. and Bersanetti,
	D. and Bertolini, A. and Betzwieser, J. and Bhagwat, S. and Bhandare,
	R. and Bilenko, I. A. and Billingsley, G. and Billman, C. R. and
	Birch, J. and Birney, I. A. and Birnholtz, O. and Biscans, S. and
	Biscoveanu, S. and Bisht, A. and Bitossi, M. and Biwer, C. and Bizouard,
	M. A. and Blackburn, J. K. and Blackman, J. and Blair, C. D. and
	Blair, D. G. and Blair, R. M. and Bloemen, S. and Bock, O. and Bode,
	N. and Boer, M. and Bogaert, G. and Bohe, A. and Bondu, F. and Bonilla,
	E. and Bonnand, R. and Boom, B. A. and Bork, R. and Boschi, V. and
	Bose, S. and Bossie, K. and Bouffanais, Y. and Bozzi, A. and Bradaschia,
	C. and Brady, P. R. and Branchesi, M. and Brau, J. E. and Briant,
	T. and Brillet, A. and Brinkmann, M. and Brisson, V. and Brockill,
	P. and Broida, J. E. and Brooks, A. F. and Brown, D. A. and Brown,
	D. D. and Brunett, S. and Buchanan, C. C. and Buikema, A. and Bulik,
	T. and Bulten, H. J. and Buonanno, A. and Buskulic, D. and Buy, C.
	and Byer, R. L. and Cabero, M. and Cadonati, L. and Cagnoli, G. and
	Cahillane, C. and Calderón Bustillo, J. and Callister, T. A. and
	Calloni, E. and Camp, J. B. and Canepa, M. and Canizares, P. and
	Cannon, K. C. and Cao, H. and Cao, J. and Capano, C. D. and Capocasa,
	E. and Carbognani, F. and Caride, S. and Carney, M. F. and Carullo,
	G. and Casanueva Diaz, J. and Casentini, C. and Caudill, S. and Cavaglià,
	M. and Cavalier, F. and Cavalieri, R. and Cella, G. and Cepeda, C. B.
	and Cerdá-Durán, P. and Cerretani, G. and Cesarini, E. and Chamberlin,
	S. J. and Chan, M. and Chao, S. and Charlton, P. and Chase, E. and
	Chassande-Mottin, E. and Chatterjee, D. and Chatziioannou, K. and
	Cheeseboro, B. D. and Chen, H. Y. and Chen, X. and Chen, Y. and Cheng,
	H.-P. and Chia, H. and Chincarini, A. and Chiummo, A. and Chmiel,
	T. and Cho, H. S. and Cho, M. and Chow, J. H. and Christensen, N.
	and Chu, Q. and Chua, A. J. K. and Chua, S. and Chung, A. K. W. and
	Chung, S. and Ciani, G. and Ciolfi, R. and Cirelli, C. E. and Cirone,
	A. and Clara, F. and Clark, J. A. and Clearwater, P. and Cleva, F.
	and Cocchieri, C. and Coccia, E. and Cohadon, P.-F. and Cohen, D.
	and Colla, A. and Collette, C. G. and Cominsky, L. R. and Constancio,
	M. and Conti, L. and Cooper, S. J. and Corban, P. and Corbitt, T. R.
	and Cordero-Carrión, I. and Corley, K. R. and Cornish, N. and Corsi,
	A. and Cortese, S. and Costa, C. A. and Coughlin, M. W. and Coughlin,
	S. B. and Coulon, J.-P. and Countryman, S. T. and Couvares, P. and
	Covas, P. B. and Cowan, E. E. and Coward, D. M. and Cowart, M. J.
	and Coyne, D. C. and Coyne, R. and Creighton, J. D. E. and Creighton,
	T. D. and Cripe, J. and Crowder, S. G. and Cullen, T. J. and Cumming,
	A. and Cunningham, L. and Cuoco, E. and Dal Canton, T. and Dálya,
	G. and Danilishin, S. L. and D’Antonio, S. and Danzmann, K. and Dasgupta,
	A. and Da Silva Costa, C. F. and Dattilo, V. and Dave, I. and Davier,
	M. and Davis, D. and Daw, E. J. and Day, B. and De, S. and DeBra,
	D. and Degallaix, J. and De Laurentis, M. and Deléglise, S. and Del
	Pozzo, W. and Demos, N. and Denker, T. and Dent, T. and De Pietri,
	R. and Dergachev, V. and De Rosa, R. and DeRosa, R. T. and De Rossi,
	C. and DeSalvo, R. and de Varona, O. and Devenson, J. and Dhurandhar,
	S. and Díaz, M. C. and Dietrich, T. and Di Fiore, L. and Di Giovanni,
	M. and Di Girolamo, T. and Di Lieto, A. and Di Pace, S. and Di Palma,
	I. and Di Renzo, F. and Doctor, Z. and Dolique, V. and Donovan, F.
	and Dooley, K. L. and Doravari, S. and Dorrington, I. and Douglas,
	R. and Dovale Álvarez, M. and Downes, T. P. and Drago, M. and Dreissigacker,
	C. and Driggers, J. C. and Du, Z. and Ducrot, M. and Dudi, R. and
	Dupej, P. and Dwyer, S. E. and Edo, T. B. and Edwards, M. C. and
	Effler, A. and Eggenstein, H.-B. and Ehrens, P. and Eichholz, J.
	and Eikenberry, S. S. and Eisenstein, R. A. and Essick, R. C. and
	Estevez, D. and Etienne, Z. B. and Etzel, T. and Evans, M. and Evans,
	T. M. and Factourovich, M. and Fafone, V. and Fair, H. and Fairhurst,
	S. and Fan, X. and Farinon, S. and Farr, B. and Farr, W. M. and Fauchon-Jones,
	E. J. and Favata, M. and Fays, M. and Fee, C. and Fehrmann, H. and
	Feicht, J. and Fejer, M. M. and Fernandez-Galiana, A. and Ferrante,
	I. and Ferreira, E. C. and Ferrini, F. and Fidecaro, F. and Finstad,
	D. and Fiori, I. and Fiorucci, D. and Fishbach, M. and Fisher, R. P.
	and Fitz-Axen, M. and Flaminio, R. and Fletcher, M. and Fong, H.
	and Font, J. A. and Forsyth, P. W. F. and Forsyth, S. S. and Fournier,
	J.-D. and Frasca, S. and Frasconi, F. and Frei, Z. and Freise, A.
	and Frey, R. and Frey, V. and Fries, E. M. and Fritschel, P. and
	Frolov, V. V. and Fulda, P. and Fyffe, M. and Gabbard, H. and Gadre,
	B. U. and Gaebel, S. M. and Gair, J. R. and Gammaitoni, L. and Ganija,
	M. R. and Gaonkar, S. G. and Garcia-Quiros, C. and Garufi, F. and
	Gateley, B. and Gaudio, S. and Gaur, G. and Gayathri, V. and Gehrels,
	N. and Gemme, G. and Genin, E. and Gennai, A. and George, D. and
	George, J. and Gergely, L. and Germain, V. and Ghonge, S. and Ghosh,
	Abhirup and Ghosh, Archisman and Ghosh, S. and Giaime, J. A. and
	Giardina, K. D. and Giazotto, A. and Gill, K. and Glover, L. and
	Goetz, E. and Goetz, R. and Gomes, S. and Goncharov, B. and González,
	G. and Gonzalez Castro, J. M. and Gopakumar, A. and Gorodetsky, M. L.
	and Gossan, S. E. and Gosselin, M. and Gouaty, R. and Grado, A. and
	Graef, C. and Granata, M. and Grant, A. and Gras, S. and Gray, C.
	and Greco, G. and Green, A. C. and Gretarsson, E. M. and Groot, P.
	and Grote, H. and Grunewald, S. and Gruning, P. and Guidi, G. M.
	and Guo, X. and Gupta, A. and Gupta, M. K. and Gushwa, K. E. and
	Gustafson, E. K. and Gustafson, R. and Halim, O. and Hall, B. R.
	and Hall, E. D. and Hamilton, E. Z. and Hammond, G. and Haney, M.
	and Hanke, M. M. and Hanks, J. and Hanna, C. and Hannam, M. D. and
	Hannuksela, O. A. and Hanson, J. and Hardwick, T. and Harms, J. and
	Harry, G. M. and Harry, I. W. and Hart, M. J. and Haster, C.-J. and
	Haughian, K. and Healy, J. and Heidmann, A. and Heintze, M. C. and
	Heitmann, H. and Hello, P. and Hemming, G. and Hendry, M. and Heng,
	I. S. and Hennig, J. and Heptonstall, A. W. and Heurs, M. and Hild,
	S. and Hinderer, T. and Ho, W. C. G. and Hoak, D. and Hofman, D.
	and Holt, K. and Holz, D. E. and Hopkins, P. and Horst, C. and Hough,
	J. and Houston, E. A. and Howell, E. J. and Hreibi, A. and Hu, Y. M.
	and Huerta, E. A. and Huet, D. and Hughey, B. and Husa, S. and Huttner,
	S. H. and Huynh-Dinh, T. and Indik, N. and Inta, R. and Intini, G.
	and Isa, H. N. and Isac, J.-M. and Isi, M. and Iyer, B. R. and Izumi,
	K. and Jacqmin, T. and Jani, K. and Jaranowski, P. and Jawahar, S.
	and Jiménez-Forteza, F. and Johnson, W. W. and Johnson-McDaniel,
	N. K. and Jones, D. I. and Jones, R. and Jonker, R. J. G. and Ju,
	L. and Junker, J. and Kalaghatgi, C. V. and Kalogera, V. and Kamai,
	B. and Kandhasamy, S. and Kang, G. and Kanner, J. B. and Kapadia,
	S. J. and Karki, S. and Karvinen, K. S. and Kasprzack, M. and Kastaun,
	W. and Katolik, M. and Katsavounidis, E. and Katzman, W. and Kaufer,
	S. and Kawabe, K. and Kéfélian, F. and Keitel, D. and Kemball, A. J.
	and Kennedy, R. and Kent, C. and Key, J. S. and Khalili, F. Y. and
	Khan, I. and Khan, S. and Khan, Z. and Khazanov, E. A. and Kijbunchoo,
	N. and Kim, Chunglee and Kim, J. C. and Kim, K. and Kim, W. and Kim,
	W. S. and Kim, Y.-M. and Kimbrell, S. J. and King, E. J. and King,
	P. J. and Kinley-Hanlon, M. and Kirchhoff, R. and Kissel, J. S. and
	Kleybolte, L. and Klimenko, S. and Knowles, T. D. and Koch, P. and
	Koehlenbeck, S. M. and Koley, S. and Kondrashov, V. and Kontos, A.
	and Korobko, M. and Korth, W. Z. and Kowalska, I. and Kozak, D. B.
	and Krämer, C. and Kringel, V. and Krishnan, B. and Królak, A. and
	Kuehn, G. and Kumar, P. and Kumar, R. and Kumar, S. and Kuo, L. and
	Kutynia, A. and Kwang, S. and Lackey, B. D. and Lai, K. H. and Landry,
	M. and Lang, R. N. and Lange, J. and Lantz, B. and Lanza, R. K. and
	Larson, S. L. and Lartaux-Vollard, A. and Lasky, P. D. and Laxen,
	M. and Lazzarini, A. and Lazzaro, C. and Leaci, P. and Leavey, S.
	and Lee, C. H. and Lee, H. K. and Lee, H. M. and Lee, H. W. and Lee,
	K. and Lehmann, J. and Lenon, A. and Leon, E. and Leonardi, M. and
	Leroy, N. and Letendre, N. and Levin, Y. and Li, T. G. F. and Linker,
	S. D. and Littenberg, T. B. and Liu, J. and Liu, X. and Lo, R. K. L.
	and Lockerbie, N. A. and London, L. T. and Lord, J. E. and Lorenzini,
	M. and Loriette, V. and Lormand, M. and Losurdo, G. and Lough, J. D.
	and Lousto, C. O. and Lovelace, G. and Lück, H. and Lumaca, D. and
	Lundgren, A. P. and Lynch, R. and Ma, Y. and Macas, R. and Macfoy,
	S. and Machenschalk, B. and MacInnis, M. and Macleod, D. M. and Magaña
	Hernandez, I. and Magaña-Sandoval, F. and Magaña Zertuche, L. and
	Magee, R. M. and Majorana, E. and Maksimovic, I. and Man, N. and
	Mandic, V. and Mangano, V. and Mansell, G. L. and Manske, M. and
	Mantovani, M. and Marchesoni, F. and Marion, F. and Márka, S. and
	Márka, Z. and Markakis, C. and Markosyan, A. S. and Markowitz, A.
	and Maros, E. and Marquina, A. and Marsh, P. and Martelli, F. and
	Martellini, L. and Martin, I. W. and Martin, R. M. and Martynov,
	D. V. and Marx, J. N. and Mason, K. and Massera, E. and Masserot,
	A. and Massinger, T. J. and Masso-Reid, M. and Mastrogiovanni, S.
	and Matas, A. and Matichard, F. and Matone, L. and Mavalvala, N.
	and Mazumder, N. and McCarthy, R. and McClelland, D. E. and McCormick,
	S. and McCuller, L. and McGuire, S. C. and McIntyre, G. and McIver,
	J. and McManus, D. J. and McNeill, L. and McRae, T. and McWilliams,
	S. T. and Meacher, D. and Meadors, G. D. and Mehmet, M. and Meidam,
	J. and Mejuto-Villa, E. and Melatos, A. and Mendell, G. and Mercer,
	R. A. and Merilh, E. L. and Merzougui, M. and Meshkov, S. and Messenger,
	C. and Messick, C. and Metzdorff, R. and Meyers, P. M. and Miao,
	H. and Michel, C. and Middleton, H. and Mikhailov, E. E. and Milano,
	L. and Miller, A. L. and Miller, B. B. and Miller, J. and Millhouse,
	M. and Milovich-Goff, M. C. and Minazzoli, O. and Minenkov, Y. and
	Ming, J. and Mishra, C. and Mitra, S. and Mitrofanov, V. P. and Mitselmakher,
	G. and Mittleman, R. and Moffa, D. and Moggi, A. and Mogushi, K.
	and Mohan, M. and Mohapatra, S. R. P. and Molina, I. and Montani,
	M. and Moore, C. J. and Moraru, D. and Moreno, G. and Morisaki, S.
	and Morriss, S. R. and Mours, B. and Mow-Lowry, C. M. and Mueller,
	G. and Muir, A. W. and Mukherjee, Arunava and Mukherjee, D. and Mukherjee,
	S. and Mukund, N. and Mullavey, A. and Munch, J. and Muñiz, E. A.
	and Muratore, M. and Murray, P. G. and Nagar, A. and Napier, K. and
	Nardecchia, I. and Naticchioni, L. and Nayak, R. K. and Neilson,
	J. and Nelemans, G. and Nelson, T. J. N. and Nery, M. and Neunzert,
	A. and Nevin, L. and Newport, J. M. and Newton, G. and Ng, K. K. Y.
	and Nguyen, P. and Nguyen, T. T. and Nichols, D. and Nielsen, A. B.
	and Nissanke, S. and Nitz, A. and Noack, A. and Nocera, F. and Nolting,
	D. and North, C. and Nuttall, L. K. and Oberling, J. and O’Dea, G. D.
	and Ogin, G. H. and Oh, J. J. and Oh, S. H. and Ohme, F. and Okada,
	M. A. and Oliver, M. and Oppermann, P. and Oram, Richard J. and O’Reilly,
	B. and Ormiston, R. and Ortega, L. F. and O’Shaughnessy, R. and Ossokine,
	S. and Ottaway, D. J. and Overmier, H. and Owen, B. J. and Pace,
	A. E. and Page, J. and Page, M. A. and Pai, A. and Pai, S. A. and
	Palamos, J. R. and Palashov, O. and Palomba, C. and Pal-Singh, A.
	and Pan, Howard and Pan, Huang-Wei and Pang, B. and Pang, P. T. H.
	and Pankow, C. and Pannarale, F. and Pant, B. C. and Paoletti, F.
	and Paoli, A. and Papa, M. A. and Parida, A. and Parker, W. and Pascucci,
	D. and Pasqualetti, A. and Passaquieti, R. and Passuello, D. and
	Patil, M. and Patricelli, B. and Pearlstone, B. L. and Pedraza, M.
	and Pedurand, R. and Pekowsky, L. and Pele, A. and Penn, S. and Perez,
	C. J. and Perreca, A. and Perri, L. M. and Pfeiffer, H. P. and Phelps,
	M. and Piccinni, O. J. and Pichot, M. and Piergiovanni, F. and Pierro,
	V. and Pillant, G. and Pinard, L. and Pinto, I. M. and Pirello, M.
	and Pitkin, M. and Poe, M. and Poggiani, R. and Popolizio, P. and
	Porter, E. K. and Post, A. and Powell, J. and Prasad, J. and Pratt,
	J. W. W. and Pratten, G. and Predoi, V. and Prestegard, T. and Prijatelj,
	M. and Principe, M. and Privitera, S. and Prix, R. and Prodi, G. A.
	and Prokhorov, L. G. and Puncken, O. and Punturo, M. and Puppo, P.
	and Pürrer, M. and Qi, H. and Quetschke, V. and Quintero, E. A. and
	Quitzow-James, R. and Raab, F. J. and Rabeling, D. S. and Radkins,
	H. and Raffai, P. and Raja, S. and Rajan, C. and Rajbhandari, B.
	and Rakhmanov, M. and Ramirez, K. E. and Ramos-Buades, A. and Rapagnani,
	P. and Raymond, V. and Razzano, M. and Read, J. and Regimbau, T.
	and Rei, L. and Reid, S. and Reitze, D. H. and Ren, W. and Reyes,
	S. D. and Ricci, F. and Ricker, P. M. and Rieger, S. and Riles, K.
	and Rizzo, M. and Robertson, N. A. and Robie, R. and Robinet, F.
	and Rocchi, A. and Rolland, L. and Rollins, J. G. and Roma, V. J.
	and Romano, J. D. and Romano, R. and Romel, C. L. and Romie, J. H.
	and Rosińska, D. and Ross, M. P. and Rowan, S. and Rüdiger, A. and
	Ruggi, P. and Rutins, G. and Ryan, K. and Sachdev, S. and Sadecki,
	T. and Sadeghian, L. and Sakellariadou, M. and Salconi, L. and Saleem,
	M. and Salemi, F. and Samajdar, A. and Sammut, L. and Sampson, L. M.
	and Sanchez, E. J. and Sanchez, L. E. and Sanchis-Gual, N. and Sandberg,
	V. and Sanders, J. R. and Sassolas, B. and Sathyaprakash, B. S. and
	Saulson, P. R. and Sauter, O. and Savage, R. L. and Sawadsky, A.
	and Schale, P. and Scheel, M. and Scheuer, J. and Schmidt, J. and
	Schmidt, P. and Schnabel, R. and Schofield, R. M. S. and Schönbeck,
	A. and Schreiber, E. and Schuette, D. and Schulte, B. W. and Schutz,
	B. F. and Schwalbe, S. G. and Scott, J. and Scott, S. M. and Seidel,
	E. and Sellers, D. and Sengupta, A. S. and Sentenac, D. and Sequino,
	V. and Sergeev, A. and Shaddock, D. A. and Shaffer, T. J. and Shah,
	A. A. and Shahriar, M. S. and Shaner, M. B. and Shao, L. and Shapiro,
	B. and Shawhan, P. and Sheperd, A. and Shoemaker, D. H. and Shoemaker,
	D. M. and Siellez, K. and Siemens, X. and Sieniawska, M. and Sigg,
	D. and Silva, A. D. and Singer, L. P. and Singh, A. and Singhal,
	A. and Sintes, A. M. and Slagmolen, B. J. J. and Smith, B. and Smith,
	J. R. and Smith, R. J. E. and Somala, S. and Son, E. J. and Sonnenberg,
	J. A. and Sorazu, B. and Sorrentino, F. and Souradeep, T. and Spencer,
	A. P. and Srivastava, A. K. and Staats, K. and Staley, A. and Steinke,
	M. and Steinlechner, J. and Steinlechner, S. and Steinmeyer, D. and
	Stevenson, S. P. and Stone, R. and Stops, D. J. and Strain, K. A.
	and Stratta, G. and Strigin, S. E. and Strunk, A. and Sturani, R.
	and Stuver, A. L. and Summerscales, T. Z. and Sun, L. and Sunil,
	S. and Suresh, J. and Sutton, P. J. and Swinkels, B. L. and Szczepańczyk,
	M. J. and Tacca, M. and Tait, S. C. and Talbot, C. and Talukder,
	D. and Tanner, D. B. and Tápai, M. and Taracchini, A. and Tasson,
	J. D. and Taylor, J. A. and Taylor, R. and Tewari, S. V. and Theeg,
	T. and Thies, F. and Thomas, E. G. and Thomas, M. and Thomas, P.
	and Thorne, K. A. and Thorne, K. S. and Thrane, E. and Tiwari, S.
	and Tiwari, V. and Tokmakov, K. V. and Toland, K. and Tonelli, M.
	and Tornasi, Z. and Torres-Forné, A. and Torrie, C. I. and Töyrä,
	D. and Travasso, F. and Traylor, G. and Trinastic, J. and Tringali,
	M. C. and Trozzo, L. and Tsang, K. W. and Tse, M. and Tso, R. and
	Tsukada, L. and Tsuna, D. and Tuyenbayev, D. and Ueno, K. and Ugolini,
	D. and Unnikrishnan, C. S. and Urban, A. L. and Usman, S. A. and
	Vahlbruch, H. and Vajente, G. and Valdes, G. and Vallisneri, M. and
	van Bakel, N. and van Beuzekom, M. and van den Brand, J. F. J. and
	Van Den Broeck, C. and Vander-Hyde, D. C. and van der Schaaf, L.
	and van Heijningen, J. V. and van Veggel, A. A. and Vardaro, M. and
	Varma, V. and Vass, S. and Vasúth, M. and Vecchio, A. and Vedovato,
	G. and Veitch, J. and Veitch, P. J. and Venkateswara, K. and Venugopalan,
	G. and Verkindt, D. and Vetrano, F. and Viceré, A. and Viets, A. D.
	and Vinciguerra, S. and Vine, D. J. and Vinet, J.-Y. and Vitale,
	S. and Vo, T. and Vocca, H. and Vorvick, C. and Vyatchanin, S. P.
	and Wade, A. R. and Wade, L. E. and Wade, M. and Walet, R. and Walker,
	M. and Wallace, L. and Walsh, S. and Wang, G. and Wang, H. and Wang,
	J. Z. and Wang, W. H. and Wang, Y. F. and Ward, R. L. and Warner,
	J. and Was, M. and Watchi, J. and Weaver, B. and Wei, L.-W. and Weinert,
	M. and Weinstein, A. J. and Weiss, R. and Wen, L. and Wessel, E. K.
	and Weßels, P. and Westerweck, J. and Westphal, T. and Wette, K.
	and Whelan, J. T. and Whitcomb, S. E. and Whiting, B. F. and Whittle,
	C. and Wilken, D. and Williams, D. and Williams, R. D. and Williamson,
	A. R. and Willis, J. L. and Willke, B. and Wimmer, M. H. and Winkler,
	W. and Wipf, C. C. and Wittel, H. and Woan, G. and Woehler, J. and
	Wofford, J. and Wong, K. W. K. and Worden, J. and Wright, J. L. and
	Wu, D. S. and Wysocki, D. M. and Xiao, S. and Yamamoto, H. and Yancey,
	C. C. and Yang, L. and Yap, M. J. and Yazback, M. and Yu, Hang and
	Yu, Haocun and Yvert, M. and Zadrożny, A. and Zanolin, M. and Zelenova,
	T. and Zendri, J.-P. and Zevin, M. and Zhang, L. and Zhang, M. and
	Zhang, T. and Zhang, Y.-H. and Zhao, C. and Zhou, M. and Zhou, Z.
	and Zhu, S. J. and Zhu, X. J. and Zimmerman, A. B. and Zucker, M. E.
	and Zweizig, J.},
  title = {GW170817: Observation of Gravitational Waves from a Binary Neutron
	Star Inspiral},
  journal = {Physical Review Letters},
  year = {2017},
  volume = {119},
  number = {16},
  month = oct,
  doi = {10.1103/physrevlett.119.161101},
  issn = {1079-7114},
  publisher = {American Physical Society (APS)},
  url = {http://dx.doi.org/10.1103/PhysRevLett.119.161101}
}

@ARTICLE{Acernese_2015_CQG,
  author = {Acernese, F and Agathos, M and Agatsuma, K and Aisa, D and Allemandou,
	N and Allocca, A and Amarni, J and Astone, P and Balestri, G and
	Ballardin, G and Barone, F and Baronick, J-P and Barsuglia, M and
	Basti, A and Basti, F and Bauer, Th S and Bavigadda, V and Bejger,
	M and Beker, M G and Belczynski, C and Bersanetti, D and Bertolini,
	A and Bitossi, M and Bizouard, M A and Bloemen, S and Blom, M and
	Boer, M and Bogaert, G and Bondi, D and Bondu, F and Bonelli, L and
	Bonnand, R and Boschi, V and Bosi, L and Bouedo, T and Bradaschia,
	C and Branchesi, M and Briant, T and Brillet, A and Brisson, V and
	Bulik, T and Bulten, H J and Buskulic, D and Buy, C and Cagnoli,
	G and Calloni, E and Campeggi, C and Canuel, B and Carbognani, F
	and Cavalier, F and Cavalieri, R and Cella, G and Cesarini, E and
	Mottin, E Chassande- and Chincarini, A and Chiummo, A and Chua, S
	and Cleva, F and Coccia, E and Cohadon, P-F and Colla, A and Colombini,
	M and Conte, A and Coulon, J-P and Cuoco, E and Dalmaz, A and D’Antonio,
	S and Dattilo, V and Davier, M and Day, R and Debreczeni, G and Degallaix,
	J and Deléglise, S and Pozzo, W Del and Dereli, H and Rosa, R De
	and Fiore, L Di and Lieto, A Di and Virgilio, A Di and Doets, M and
	Dolique, V and Drago, M and Ducrot, M and Endrőczi, G and Fafone,
	V and Farinon, S and Ferrante, I and Ferrini, F and Fidecaro, F and
	Fiori, I and Flaminio, R and Fournier, J-D and Franco, S and Frasca,
	S and Frasconi, F and Gammaitoni, L and Garufi, F and Gaspard, M
	and Gatto, A and Gemme, G and Gendre, B and Genin, E and Gennai,
	A and Ghosh, S and Giacobone, L and Giazotto, A and Gouaty, R and
	Granata, M and Greco, G and Groot, P and Guidi, G M and Harms, J
	and Heidmann, A and Heitmann, H and Hello, P and Hemming, G and Hennes,
	E and Hofman, D and Jaranowski, P and Jonker, R J G and Kasprzack,
	M and Kéfélian, F and Kowalska, I and Kraan, M and Królak, A and
	Kutynia, A and Lazzaro, C and Leonardi, M and Leroy, N and Letendre,
	N and Li, T G F and Lieunard, B and Lorenzini, M and Loriette, V
	and Losurdo, G and Magazzù, C and Majorana, E and Maksimovic, I and
	Malvezzi, V and Man, N and Mangano, V and Mantovani, M and Marchesoni,
	F and Marion, F and Marque, J and Martelli, F and Martellini, L and
	Masserot, A and Meacher, D and Meidam, J and Mezzani, F and Michel,
	C and Milano, L and Minenkov, Y and Moggi, A and Mohan, M and Montani,
	M and Morgado, N and Mours, B and Mul, F and Nagy, M F and Nardecchia,
	I and Naticchioni, L and Nelemans, G and Neri, I and Neri, M and
	Nocera, F and Pacaud, E and Palomba, C and Paoletti, F and Paoli,
	A and Pasqualetti, A and Passaquieti, R and Passuello, D and Perciballi,
	M and Petit, S and Pichot, M and Piergiovanni, F and Pillant, G and
	Piluso, A and Pinard, L and Poggiani, R and Prijatelj, M and Prodi,
	G A and Punturo, M and Puppo, P and Rabeling, D S and Rácz, I and
	Rapagnani, P and Razzano, M and Re, V and Regimbau, T and Ricci,
	F and Robinet, F and Rocchi, A and Rolland, L and Romano, R and Rosińska,
	D and Ruggi, P and Saracco, E and Sassolas, B and Schimmel, F and
	Sentenac, D and Sequino, V and Shah, S and Siellez, K and Straniero,
	N and Swinkels, B and Tacca, M and Tonelli, M and Travasso, F and
	Turconi, M and Vajente, G and van Bakel, N and van Beuzekom, M and
	van den Brand, J F J and Van Den Broeck, C and van der Sluys, M V
	and van Heijningen, J and Vasúth, M and Vedovato, G and Veitch, J
	and Verkindt, D and Vetrano, F and Viceré, A and Vinet, J-Y and Visser,
	G and Vocca, H and Ward, R and Was, M and Wei, L-W and Yvert, M and
	żny, A Zadro and Zendri, J-P},
  title = {Advanced Virgo: a second-generation interferometric gravitational
	wave detector},
  journal = {Classical and Quantum Gravity},
  year = {2015},
  volume = {32},
  pages = {024001},
  number = {2},
  month = {dec},
  doi = {10.1088/0264-9381/32/2/024001},
  publisher = {IOP Publishing},
  url = {https://doi.org/10.1088/0264-9381/32/2/024001}
}

@ARTICLE{Akutsu_2019_NA,
  author = {Akutsu, T. and Ando, M. and Arai, K. and Arai, Y. and Araki, Sakae
	and Araya, Akito and Aritomi, N. and Asada, Hideki and Aso, Yoichi
	and Atsuta, S. and Awai, K. and Bae, Susung and Baiotti, Luca and
	Barton, M. and Cannon, Kipp and Capocasa, E. and Chen, C-S and Chiu,
	Ting-Wai and Cho, Kyuman and collaboration, KAGRA},
  title = {KAGRA: 2.5 generation interferometric gravitational wave detector},
  journal = {Nature Astronomy},
  year = {2019},
  volume = {3},
  pages = {35-40},
  month = {01},
  doi = {10.1038/s41550-018-0658-y}
}

@ARTICLE{Arzoumanian_2018_APJSS,
  author = {Arzoumanian, Zaven and Brazier, Adam and Burke-Spolaor, Sarah and
	Chamberlin, Sydney and Chatterjee, Shami and Christy, Brian and Cordes,
	James M. and Cornish, Neil J. and Crawford, Fronefield and Cromartie,
	H. Thankful and Crowter, Kathryn and DeCesar, Megan E. and Demorest,
	Paul B. and Dolch, Timothy and Ellis, Justin A. and Ferdman, Robert
	D. and Ferrara, Elizabeth C. and Fonseca, Emmanuel and Garver-Daniels,
	Nathan and Gentile, Peter A. and Halmrast, Daniel and Huerta, E.
	A. and Jenet, Fredrick A. and Jessup, Cody and Jones, Glenn and Jones,
	Megan L. and Kaplan, David L. and Lam, Michael T. and W. Lazio, T.
	Joseph and Levin, Lina and Lommen, Andrea and Lorimer, Duncan R.
	and Luo, Jing and Lynch, Ryan S. and Madison, Dustin and Matthews,
	Allison M. and McLaughlin, Maura A. and McWilliams, Sean T. and Mingarelli,
	Chiara and Ng, Cherry and Nice, David J. and Pennucci, Timothy T.
	and Ransom, Scott M. and Ray, Paul S. and Siemens, Xavier and Simon,
	Joseph and Spiewak, Renée and Stairs, Ingrid H. and Stinebring, Daniel
	R. and Stovall, Kevin and Swiggum, Joseph K. and Taylor, Stephen
	R. and Vallisneri, Michele and van Haasteren, Rutger and Vigeland,
	Sarah J. and Zhu, Weiwei},
  title = {The NANOGrav 11-year Data Set: High-precision Timing of 45 Millisecond
	Pulsars},
  journal = {The Astrophysical Journal Supplement Series},
  year = {2018},
  volume = {235},
  pages = {37},
  number = {2},
  month = apr,
  doi = {10.3847/1538-4365/aab5b0},
  issn = {1538-4365},
  publisher = {American Astronomical Society},
  url = {http://dx.doi.org/10.3847/1538-4365/aab5b0}
}

@ARTICLE{Baldo_2016_PPNP,
  author = {M. Baldo and G.F. Burgio},
  title = {The nuclear symmetry energy},
  journal = {Progress in Particle and Nuclear Physics},
  year = {2016},
  volume = {91},
  pages = {203-258},
  doi = {https://doi.org/10.1016/j.ppnp.2016.06.006},
  issn = {0146-6410},
  url = {https://www.sciencedirect.com/science/article/pii/S0146641016300254}
}

@ARTICLE{Bombaci_1991_PRC,
  author = {Bombaci, Ignazio and Lombardo, U.},
  title = {Asymmetric nuclear matter equation of state},
  journal = {Physical review C: Nuclear physics},
  year = {1991},
  volume = {44},
  pages = {1892-1900},
  month = {12},
  doi = {10.1103/PhysRevC.44.1892}
}

@ARTICLE{Brockmann_1990_PRC,
  author = {Brockmann, R. and Machleidt, R.},
  title = {Relativistic nuclear structure. I. Nuclear matter},
  journal = {Phys. Rev. C},
  year = {1990},
  volume = {42},
  pages = {1965--1980},
  month = {Nov},
  doi = {10.1103/PhysRevC.42.1965},
  issue = {5},
  numpages = {0},
  publisher = {American Physical Society},
  url = {https://link.aps.org/doi/10.1103/PhysRevC.42.1965}
}

@article{2024-Machleidt-PPNP,
title = {Recent advances in chiral EFT based nuclear forces and their applications},
journal = {Progress in Particle and Nuclear Physics},
volume = {137},
pages = {104117},
year = {2024},
issn = {0146-6410},
doi = {https://doi.org/10.1016/j.ppnp.2024.104117},
url = {https://www.sciencedirect.com/science/article/pii/S0146641024000218},
author = {R. Machleidt and F. Sammarruca}
}

@Article{2021-Burgio-PPNP,
  author    = {G.F. Burgio and H.-J. Schulze and I. Vidaña and J.-B. Wei},
  journal   = {Progress in Particle and Nuclear Physics},
  title     = {Neutron stars and the nuclear equation of state},
  year      = {2021},
  issn      = {0146-6410},
  pages     = {103879},
  volume    = {120},
  doi       = {https://doi.org/10.1016/j.ppnp.2021.103879},
  url       = {https://www.sciencedirect.com/science/article/pii/S0146641021000338},
}

@ARTICLE{Choudhury_2024_APJL,
  author = {Choudhury, Devarshi and Salmi, Tuomo and Vinciguerra, Serena and
	Riley, Thomas E. and Kini, Yves and Watts, Anna L. and Dorsman, Bas
	and Bogdanov, Slavko and Guillot, Sebastien and Ray, Paul S. and
	Reardon, Daniel J. and Remillard, Ronald A. and Bilous, Anna V. and
	Huppenkothen, Daniela and Lattimer, James M. and Rutherford, Nathan
	and Arzoumanian, Zaven and Gendreau, Keith C. and Morsink, Sharon
	M. and Ho, Wynn C. G.},
  title = {A NICER View of the Nearest and Brightest Millisecond Pulsar: PSR
	J0437–4715},
  journal = {The Astrophysical Journal Letters},
  year = {2024},
  volume = {971},
  pages = {L20},
  number = {1},
  month = aug,
  doi = {10.3847/2041-8213/ad5a6f},
  issn = {2041-8213},
  publisher = {American Astronomical Society},
  url = {http://dx.doi.org/10.3847/2041-8213/ad5a6f}
}

@article{2024-Sorensen-PPNP,
title = {Dense nuclear matter equation of state from heavy-ion collisions},
journal = {Progress in Particle and Nuclear Physics},
volume = {134},
pages = {104080},
year = {2024},
issn = {0146-6410},
doi = {https://doi.org/10.1016/j.ppnp.2023.104080},
url = {https://www.sciencedirect.com/science/article/pii/S0146641023000613},
author = {Agnieszka Sorensen and Kshitij Agarwal and Kyle W. Brown and Zbigniew Chajęcki and Paweł Danielewicz and Christian Drischler and Stefano Gandolfi and Jeremy W. Holt and Matthias Kaminski and Che-Ming Ko and Rohit Kumar and Bao-An Li and William G. Lynch and Alan B. McIntosh and William G. Newton and Scott Pratt and Oleh Savchuk and Maria Stefaniak and Ingo Tews and ManYee Betty Tsang and Ramona Vogt and Hermann Wolter and Hanna Zbroszczyk and Navid Abbasi and Jörg Aichelin and Anton Andronic and Steffen A. Bass and Francesco Becattini and David Blaschke and Marcus Bleicher and Christoph Blume and Elena Bratkovskaya and B. Alex Brown and David A. Brown and Alberto Camaiani and Giovanni Casini and Katerina Chatziioannou and Abdelouahad Chbihi and Maria Colonna and Mircea Dan Cozma and Veronica Dexheimer and Xin Dong and Travis Dore and Lipei Du and José A. Dueñas and Hannah Elfner and Wojciech Florkowski and Yuki Fujimoto and Richard J. Furnstahl and Alexandra Gade and Tetyana Galatyuk and Charles Gale and Frank Geurts and Fabiana Gramegna and Sašo Grozdanov and Kris Hagel and Steven P. Harris and Wick Haxton and Ulrich Heinz and Michal P. Heller and Or Hen and Heiko Hergert and Norbert Herrmann and Huan Zhong Huang and Xu-Guang Huang and Natsumi Ikeno and Gabriele Inghirami and Jakub Jankowski and Jiangyong Jia and José C. Jiménez and Joseph Kapusta and Behruz Kardan and Iurii Karpenko and Declan Keane and Dmitri Kharzeev and Andrej Kugler and Arnaud {Le Fèvre} and Dean Lee and Hong Liu and Michael A. Lisa and William J. Llope and Ivano Lombardo and Manuel Lorenz and Tommaso Marchi and Larry McLerran and Ulrich Mosel and Anton Motornenko and Berndt Müller and Paolo Napolitani and Joseph B. Natowitz and Witold Nazarewicz and Jorge Noronha and Jacquelyn Noronha-Hostler and Grażyna Odyniec and Panagiota Papakonstantinou and Zuzana Paulínyová and Jorge Piekarewicz and Robert D. Pisarski and Christopher Plumberg and Madappa Prakash and Jørgen Randrup and Claudia Ratti and Peter Rau and Sanjay Reddy and Hans-Rudolf Schmidt and Paolo Russotto and Radoslaw Ryblewski and Andreas Schäfer and Björn Schenke and Srimoyee Sen and Peter Senger and Richard Seto and Chun Shen and Bradley Sherrill and Mayank Singh and Vladimir Skokov and Michał Spaliński and Jan Steinheimer and Mikhail Stephanov and Joachim Stroth and Christian Sturm and Kai-Jia Sun and Aihong Tang and Giorgio Torrieri and Wolfgang Trautmann and Giuseppe Verde and Volodymyr Vovchenko and Ryoichi Wada and Fuqiang Wang and Gang Wang and Klaus Werner and Nu Xu and Zhangbu Xu and Ho-Ung Yee and Sherry Yennello and Yi Yin}
}

@ARTICLE{Van2005PRL,
  author = {van Dalen, E. N. E. and Fuchs, C. and Faessler, Amand},
  title = {Effective Nucleon Masses in Symmetric and Asymmetric Nuclear Matter},
  journal = {Physical Review Letters},
  year = {2005},
  volume = {95},
  pages = {022302},
  number = {2},
  doi = {10.1103/PhysRevLett.95.022302},
  type = {Journal Article},
  url = {https://link.aps.org/doi/10.1103/PhysRevLett.95.022302}
}

@ARTICLE{Pawel_2002_Science,
  author = {Paweł Danielewicz and Roy Lacey and William G. Lynch},
  title = {Determination of the Equation of State of Dense Matter},
  journal = {Science},
  year = {2002},
  volume = {298},
  pages = {1592-1596},
  number = {5598},
  doi = {10.1126/science.1078070},
  eprint = {https://www.science.org/doi/pdf/10.1126/science.1078070},
  url = {https://www.science.org/doi/abs/10.1126/science.1078070}
}

@ARTICLE{Demorest_2010_Nature,
  author = {Demorest, P. B. and Pennucci, T. and Ransom, S. M. and Roberts, M.
	S. E. and Hessels, J. W. T.},
  title = {A two-solar-mass neutron star measured using Shapiro delay},
  journal = {Nature},
  year = {2010},
  volume = {467},
  pages = {1081–1083},
  number = {7319},
  month = oct,
  doi = {10.1038/nature09466},
  issn = {1476-4687},
  publisher = {Springer Science and Business Media LLC},
  url = {http://dx.doi.org/10.1038/nature09466}
}

@ARTICLE{Tim.Dietrich_2020_Science,
  author = {Tim Dietrich and Michael W. Coughlin and Peter T. H. Pang and Mattia
	Bulla and Jack Heinzel and Lina Issa and Ingo Tews and Sarah Antier},
  title = {Multimessenger constraints on the neutron-star equation of state
	and the Hubble constant},
  journal = {Science},
  year = {2020},
  volume = {370},
  pages = {1450-1453},
  number = {6523},
  doi = {10.1126/science.abb4317},
  eprint = {https://www.science.org/doi/pdf/10.1126/science.abb4317},
  url = {https://www.science.org/doi/abs/10.1126/science.abb4317}
}

@ARTICLE{Margherita_2019_PRL,
  author = {Margherita Fasano and Tiziano Abdelsalhin and Andrea Maselli and
	Valeria Ferrari},
  title = {Constraining the Neutron Star Equation of State Using Multiband Independent
	Measurements of Radii and Tidal Deformabilities.},
  journal = {Physical review letters},
  year = {2019},
  volume = {123 14},
  pages = {141101},
  url = {https://api.semanticscholar.org/CorpusID:119079454}
}

@ARTICLE{Ferreira_2021_JCAP,
  author = {Ferreira, Márcio and Providência, Constança},
  title = {Unveiling the nuclear matter EoS from neutron star properties: a
	supervised machine learning approach},
  journal = {Journal of Cosmology and Astroparticle Physics},
  year = {2021},
  volume = {2021},
  pages = {011},
  month = {07},
  doi = {10.1088/1475-7516/2021/07/011}
}

@ARTICLE{Fonseca_2016_APJ,
  author = {Fonseca, Emmanuel and Pennucci, Timothy T. and Ellis, Justin A. and
	Stairs, Ingrid H. and Nice, David J. and Ransom, Scott M. and Demorest,
	Paul B. and Arzoumanian, Zaven and Crowter, Kathryn and Dolch, Timothy
	and Ferdman, Robert D. and Gonzalez, Marjorie E. and Jones, Glenn
	and Jones, Megan L. and Lam, Michael T. and Levin, Lina and McLaughlin,
	Maura A. and Stovall, Kevin and Swiggum, Joseph K. and Zhu, Weiwei},
  title = {THE NANOGRAV NINE-YEAR DATA SET: MASS AND GEOMETRIC MEASUREMENTS
	OF BINARY MILLISECOND PULSARS},
  journal = {The Astrophysical Journal},
  year = {2016},
  volume = {832},
  pages = {167},
  number = {2},
  month = nov,
  doi = {10.3847/0004-637x/832/2/167},
  issn = {1538-4357},
  publisher = {American Astronomical Society},
  url = {http://dx.doi.org/10.3847/0004-637X/832/2/167}
}

@ARTICLE{Fujimoto_2024_PRD,
  author = {Fujimoto, Yuki and Fukushima, Kenji and Kamata, Syo and Murase, Koichi},
  title = {Uncertainty quantification in the machine-learning inference from
	neutron star probability distribution to the equation of state},
  journal = {Physical Review D},
  year = {2024},
  volume = {110},
  month = {08},
  doi = {10.1103/PhysRevD.110.034035}
}

@ARTICLE{Fujimoto_2021_JHEP,
  author = {Fujimoto, Yuki and Fukushima, Kenji and Murase, Koichi},
  title = {Extensive studies of the neutron star equation of state from the
	deep learning inference with the observational data augmentation},
  journal = {Journal of High Energy Physics},
  year = {2021},
  volume = {2021},
  month = {03},
  doi = {10.1007/JHEP03(2021)273}
}

@ARTICLE{Fujimoto_2020_PRD,
  author = {Fujimoto, Yuki and Fukushima, Kenji and Murase, Koichi},
  title = {Mapping neutron star data to the equation of state using the deep
	neural network},
  journal = {Phys. Rev. D},
  year = {2020},
  volume = {101},
  pages = {054016},
  month = {Mar},
  doi = {10.1103/PhysRevD.101.054016},
  issue = {5},
  numpages = {6},
  publisher = {American Physical Society},
  url = {https://link.aps.org/doi/10.1103/PhysRevD.101.054016}
}

@article{2018-Fujimoto-PhysRevD.98.023019,
  title = {Methodology study of machine learning for the neutron star equation of state},
  author = {Fujimoto, Yuki and Fukushima, Kenji and Murase, Koichi},
  journal = {Phys. Rev. D},
  volume = {98},
  issue = {2},
  pages = {023019},
  numpages = {6},
  year = {2018},
  month = {Jul},
  publisher = {American Physical Society},
  doi = {10.1103/PhysRevD.98.023019},
  url = {https://link.aps.org/doi/10.1103/PhysRevD.98.023019}
}

@article{2020-YangJJ-ARNPS,
   author = "Yang, Junjie and Piekarewicz, J.",
   title = "Covariant Density Functional Theory in Nuclear Physics and Astrophysics", 
   journal= "Annual Review of Nuclear and Particle Science",
   year = "2020",
   volume = "70",
   number = "Volume 70, 2020",
   pages = "21-41",
   doi = "https://doi.org/10.1146/annurev-nucl-101918-023608",
   url = "https://www.annualreviews.org/content/journals/10.1146/annurev-nucl-101918-023608",
   publisher = "Annual Reviews",
   issn = "1545-4134",
   type = "Journal Article",
  }

@article{2023-Sedrakian-PPNP,
title = {Heavy baryons in compact stars},
journal = {Progress in Particle and Nuclear Physics},
volume = {131},
pages = {104041},
year = {2023},
issn = {0146-6410},
doi = {https://doi.org/10.1016/j.ppnp.2023.104041},
url = {https://www.sciencedirect.com/science/article/pii/S0146641023000224},
author = {Armen Sedrakian and Jia Jie Li and Fridolin Weber}
}

\end{document}